\documentclass[aip, amsmath,amssymb,reprint]{revtex4-1}
\usepackage{graphicx}
\usepackage{dcolumn}
\usepackage{bm}

\usepackage[utf8]{inputenc}
\usepackage[T1]{fontenc}
\usepackage{mathptmx}
\usepackage{etoolbox}
\usepackage{sidecap}
\usepackage{floatrow}
\usepackage[bottom]{footmisc}
\usepackage{verbatim}
\usepackage[normalem]{ulem}
\usepackage{hyperref}
\hypersetup{
    colorlinks=true,
    linkcolor=black,
    filecolor=black, 
    citecolor=black,
    urlcolor=blue,
    }

\makeatletter
\def\@email#1#2{%
 \endgroup
 \patchcmd{\titleblock@produce}
  {\frontmatter@RRAPformat}
  {\frontmatter@RRAPformat{\produce@RRAP{*#1\href{mailto:#2}{#2}}}\frontmatter@RRAPformat}
  {}{}
}%
\makeatother
\begin{document}

\preprint{AIP/123-QED}

\title{Machine Learning Predictions of High-Curie-Temperature Materials}
\author{Joshua F. Belot}%
\affiliation{ 
Brigham Young University, Provo, UT, USA, 84602
}%
 
\author{Valentin Taufour}%
\affiliation{ 
University of California, Davis, One Shields Avenue, Davis, CA, USA, 95616
}%

\author{Stefano Sanvito}
\affiliation{%
School of Physics, AMBER and CRANN Institute, Trinity College, Dublin 2, Ireland
}%

\author{Gus L. W. Hart}
    \email{gus.hart@byu.edu}
\affiliation{%
Brigham Young University, Provo, UT, USA, 84602
}%

\date{\today}

\begin{abstract}
Technologies that function at room temperature often require magnets with a high Curie temperature, $T_\mathrm{C}$, and can be improved with better materials. Discovering magnetic materials with a substantial $T_\mathrm{C}$ is challenging because of the large number of candidates and the cost of fabricating and testing them. Using the two largest known data sets of experimental Curie temperatures, we develop machine-learning models to make rapid $T_\mathrm{C}$ predictions solely based on the chemical composition of a material. We train a random forest model and a $k$-NN one and predict on an initial dataset of over 2,500 materials and then validate the model on a new dataset containing over 3,000 entries. The accuracy is compared for multiple compounds' representations (``descriptors'') and regression approaches. A random forest model provides the most accurate predictions and is not improved by dimensionality reduction or by using more complex descriptors based on atomic properties. A random forest model trained on a combination of both datasets shows that cobalt-rich and iron-rich materials have the highest Curie temperatures for all binary and ternary compounds. An analysis of the model reveals systematic error that causes the model to over-predict low-$T_\mathrm{C}$ materials and under-predict high-$T_\mathrm{C}$ materials. For exhaustive searches to find new high-$T_\mathrm{C}$ materials,  analysis of the learning rate suggests either that much more data is needed or that more efficient descriptors are necessary. 
\end{abstract}




\maketitle


Ferromagnetic materials are essential in modern technologies especially in energy production and data storage. The threshold temperature where magnetism disappears is called the Curie temperature, $T_\mathrm{C}$. Searches for high-$T_\mathrm{C}$ materials typically look for magnets with a $T_\mathrm{C}$ of at least 550--600\,K, which is required for reliably running an application at room temperature.\cite{nelson2019predicting, coey2020perspective} High-$T_\mathrm{C}$ magnets are valuable but rare,\cite{nelson2019predicting, curtarolo2013high, merker2022machine} in particular when other properties are required on their electronic structure.\cite{Roadmap2020} \textcolor{black}{Although some empirical rules for the design of new magnets exist\cite{coey2010magnetism}, rapid predictions of Curie temperatures could assist in identifying candidate high-TC magnets in large-scale screening exercises. } Thousands of known ferromagnetic materials exist~\cite{byland2022statistics}, and while high-throughput computation can help in identifying hundreds of thousands of \emph{potential} magnets, experience suggests that only a fraction of them can actually be realized.\cite{curtarolo2013high, kabiraj2020high, graf2011simple, doi:10.1126/sciadv.1602241}

Measuring the Curie temperature of a compound is a relatively standard and accurate procedure, but of course needs the material to be made first. In contrast, in a computational design process, the $T_\mathrm{C}$ must be predicted ahead of experiments, only using physical and chemical information. This is a complex task prone to large errors. In fact, one needs to compute the elementary magnetic excitations of a compound, most typically from density functional theory (DFT), map these on a simple model, usually a Heisenberg-type one, and then perform thermodynamic sampling with Monte Carlo methods.\cite{nolting1995electronic, kubler2006ab, kubler2007understanding, halilov1997magnon, kormann2008free, PhysRevB.55.14975, levzaic2007first, PhysRevB.73.214412, PhysRevB.75.054402, tanaka2020prediction, kormann2009pressure, liechtenstein1987local, pajda2001ab, turek2005electronic, gong2019calculating, takahashi2007first} Then, the choice of DFT functional, the quality and appropriateness of the mapping, and subtleties in the Monte Carlo algorithms, all contribute to a large uncertainty on the predictions. Often this uncertainty is so severe that blind predictions of $T_\mathrm{C}$ for unknown compounds are almost impossible to make. 


Machine-learning algorithms capture complex relationships in data that may be difficult to recognize or understand.\cite{song2020machine} Even though machine-learning models often have low interpretability, they can provide cheap Curie temperature predictions. Rapid predictions can be a first step in screening computer-suggested materials, so that experimental efforts can be more selective. 

\textcolor{black}{Algorithms such as ridge regression, kernel ridge regression, neural networks, and random forest have been trained on experimental data and used to predict Curie temperatures.\cite{Sanvito2017SA, nelson2019predicting, nguyen2019regression, zhai2018accelerated, long2021accelerating, kabiraj2020high, dam2018important,Hu2020JPCM, Lu2022Chem}. Most of these studies were limited to specific structural or chemical families, and used a relatively small dataset ($<1800$) for training which limited the ability to generalize the results to other less known families.
In this work, a random-forest model is trained on an initial dataset of the chemical compositions of over 2,500 ferromagnetic materials~\cite{nelson2019predicting}. We then used a larger dataset of $\sim3100$~\cite{byland2022statistics}  for a ``blind test'', which shows our model to have good generalization. Then we built a model using the combined dataset. This is, by far, the largest dataset to be used in this kind of study.}

\begin{figure*}
        \center \includegraphics[width=1.0\linewidth]{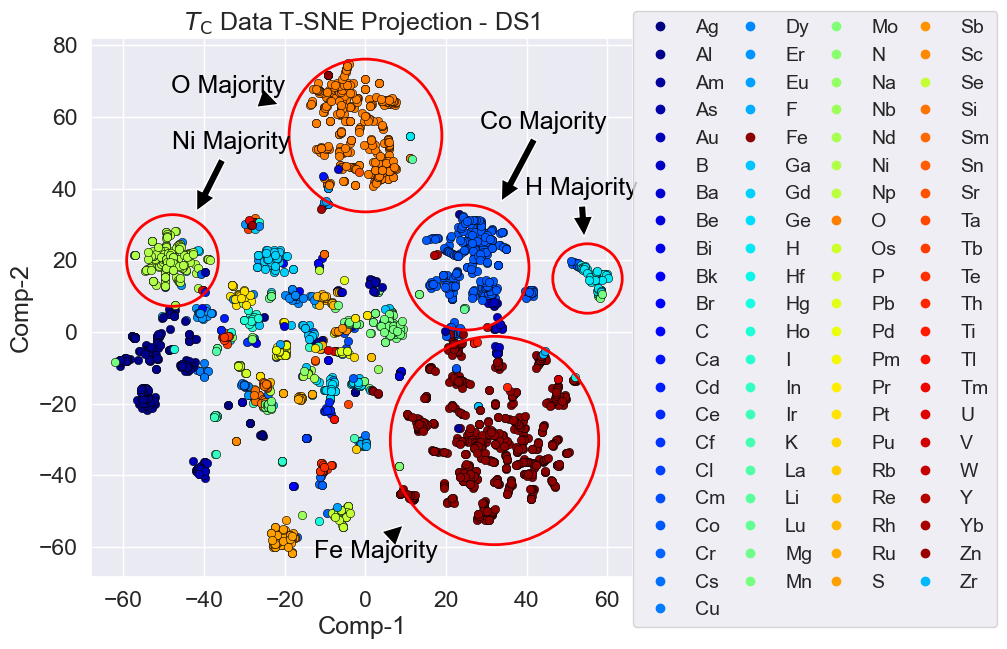}
        \label{fig:ELEMTSNE}
\end{figure*}

\begin{figure*}
    \floatbox[{\capbeside\thisfloatsetup{capbesideposition={right,center},capbesidewidth=6cm}}]{figure}[\FBwidth]
    {\caption{(a) Experimental data (DS1) projected from the 85-dimensional feature space to two dimensions in a t-SNE\cite{tsne} plot. \textcolor{black}{Colors were assigned based on the majority element in each compound.} \textcolor{black}{} (b) DS1 represented as a two-dimensional t-SNE plot. Colors were assigned based on the Curie temperature. Two compounds, Fe$_1$Ni$_3$ and Cr$_2$Pt$_3$, have anomalously high Curie temperatures, relative to others in the same cluster.}
    \label{fig:TCTSNE}}
    {\includegraphics[width=10cm]{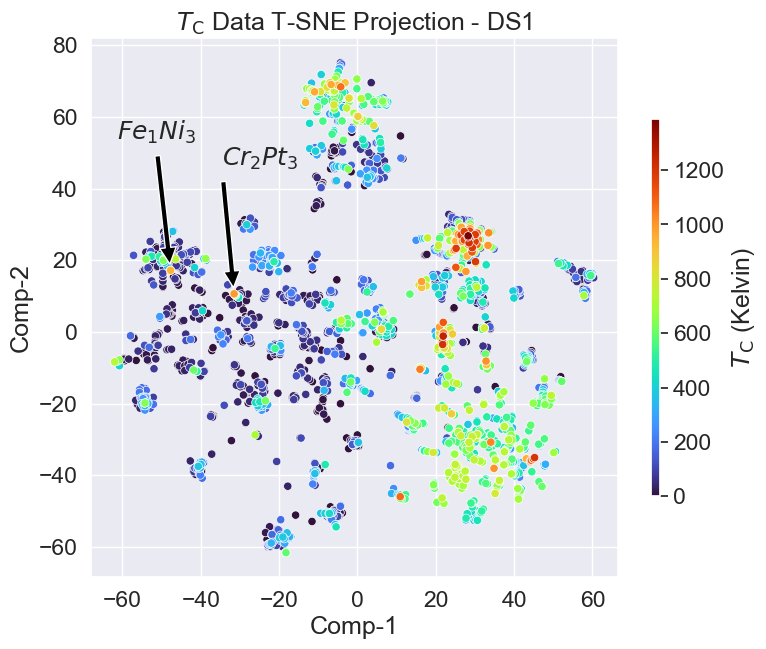}}
\end{figure*}

\begin{figure*}
        \center \includegraphics[width=1.0\linewidth]{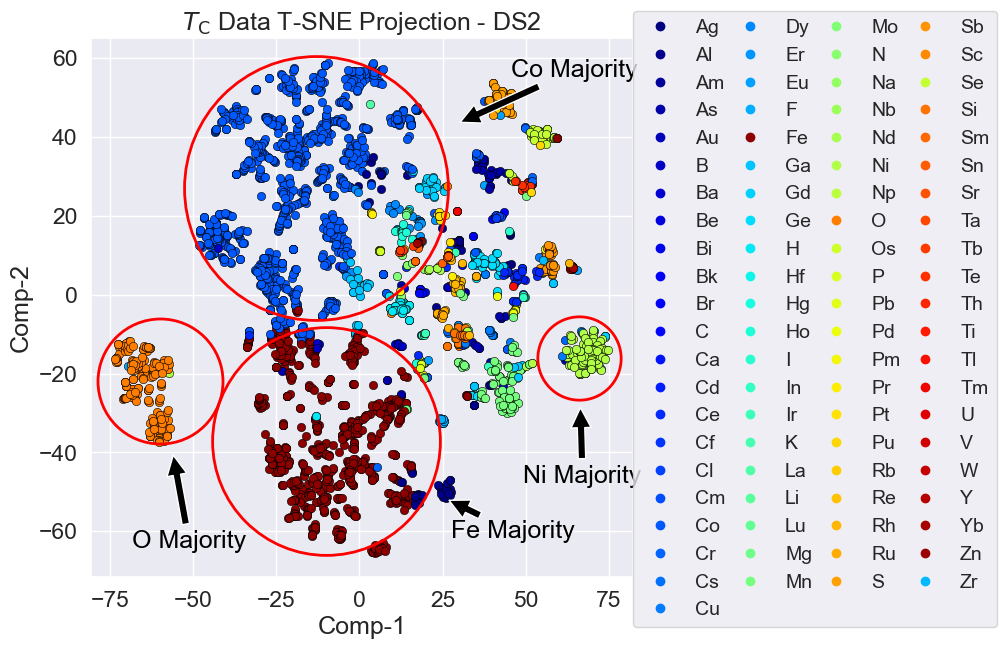}
        \label{fig:ELEMTSNE-DS2}
\end{figure*}

\begin{figure*}
    \floatbox[{\capbeside\thisfloatsetup{capbesideposition={right,center},capbesidewidth=6cm}}]{figure}[\FBwidth]
    {\caption{(a) Experimental data (DS2) projected from the 85-dimensional feature space to two dimensions in a t-SNE\cite{tsne} plot. \textcolor{black}{Colors were assigned based on the majority element in each compound.} \textcolor{black}{} (b) DS2 represented as a two-dimensional t-SNE plot. Colors were assigned based on the Curie temperature.}
    \label{fig:TCTSNE-DS2}}
    {\includegraphics[width=10cm]{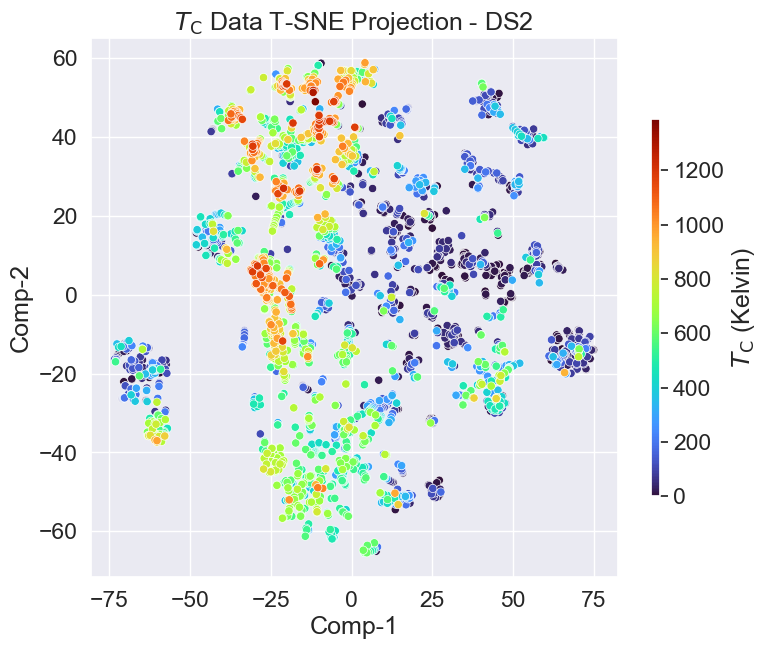}}
    
\end{figure*}

In general, the crystal structure affects the Curie temperature and there have been attempts to incorporate structural information into different machine-learning models with varying degrees of success. In some cases, Curie temperature predictions that utilize known structural data are more accurate, in particular when the structural diversity present in the dataset is limited.\cite{nguyen2019regression, long2021accelerating, balachandran2019machine, balachandran2016structure, kabiraj2020high, byland2022statistics, chen2022accurate} However, the use of structural information in a larger and more diverse set of data, may have an adverse effect on the $T_\mathrm{C}$ predictions.\cite{nelson2019predicting} Furthermore, the lack of available structural information excludes a large portion of experimental data which weakens the model's predictive power. The use of structural data in training a machine learning model also limits predictions to materials with known structure. In this work, we avoid these problems and exploit our large corpus of magnetic data in DS1 and DS2 by using models and data that do not include structural information.

In reference~[\onlinecite{nelson2019predicting}], a random forest model was used to predict the Curie temperature based solely on chemical composition. In this work, we first use the same data (DS1) and attempt to improve the predictions by trying different machine-learning models and alternate descriptors. Using the random forest model trained on DS1 to make predictions on a new set of experimental data (DS2),\cite{byland2022statistics} we find that the model generalizes quite well \textcolor{black}{but the prediction errors are 90 K (MAE) as compared to 69 K (MAE).} Next, we combine DS1 and DS2 to build a new model and scan over 100s of thousands of potential new magnets. In the process, we find that the errors between the experimental and predicted Curie temperatures reveal a previously unnoticed systematic error. 

\begin{figure*}
    \begin{minipage}[b]{0.45\linewidth}
        \centering
        \includegraphics[width=\textwidth]{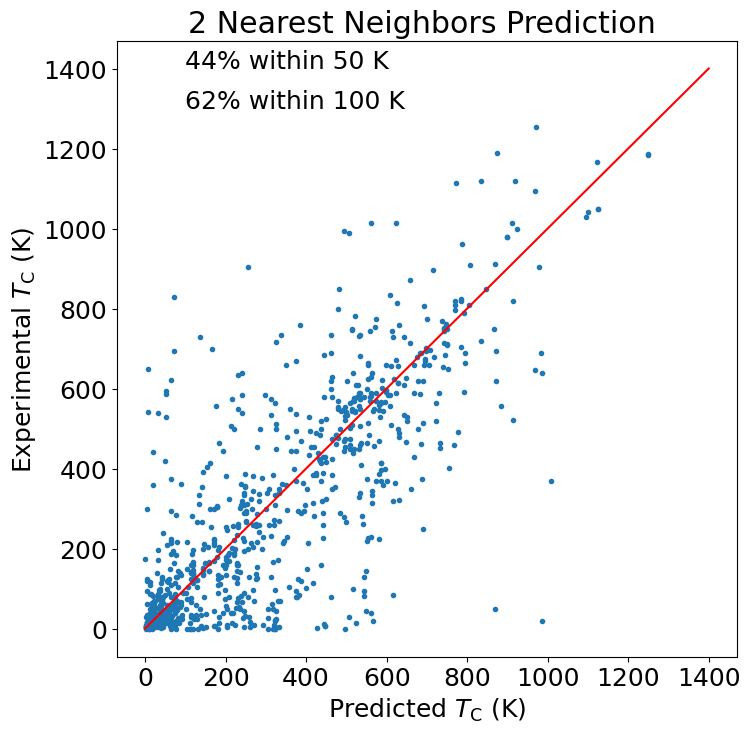}
        \label{fig:figure1}
    \end{minipage}
    \hspace{0.1cm}
    \begin{minipage}[b]{0.45\linewidth}
        \centering
        \includegraphics[width=\textwidth]{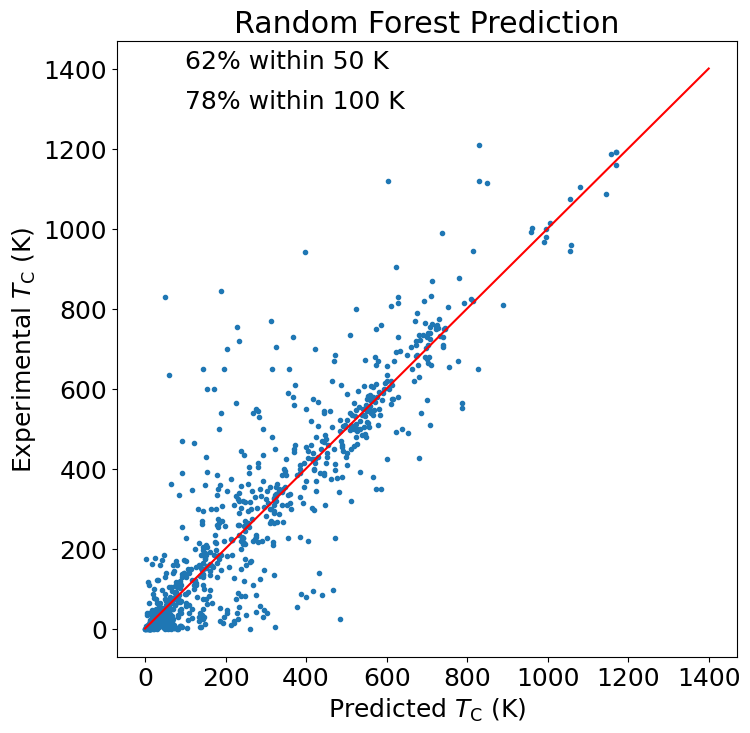}
        \label{fig:figure2}
    \end{minipage}
    \caption{$k$-nearest neighbor and random forest predictions. (Left) Prediction with $k$-NN model with $k$ = 2. 44\% of the $T_\mathrm{C}$ predictions were within 50 K of the actual values and 64\% were within 100 K. The mean absolute error was 109 K \textcolor{black}{(A 2-NN model with randomly shuffled $T_\mathrm{C}$ values gives a MAE of 270 K)}. (Right) Random forest prediction. 62\% of the $T_\mathrm{C}$ predictions were within 50 K of the actual values and 78\% were within 100 K. The mean absolute error was 69 K. \textcolor{black}{(A random forest model with randomly shuffled $T_\mathrm{C}$ values gives a MAE of 240 K)}.
    \label{fig:NSdataPredictions}}
    
\end{figure*}

\begin{figure}
    \caption{Random forest model trained on DS1 and validated with DS2. 54\% of the $T_\mathrm{C}$ predictions for DS2 were within 50 K of the actual values and 70\% were within 100 K. The mean absolute error was 90 K. \textcolor{black}{}}
    \includegraphics[width=8.8cm]{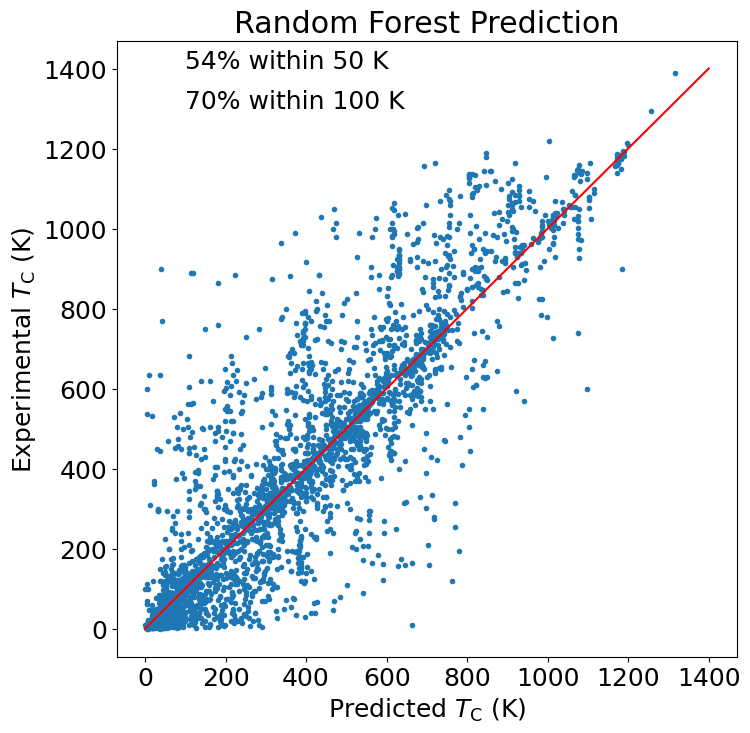}
    \label{fig:Random Forest DS2}
\end{figure}

We cleaned both DS1 and DS2 according to the methods outlined
in Ref.~[\onlinecite{nelson2019predicting}]. Both datasets had many duplicate compounds often with different reported experimental Curie temperatures. The duplicates were eliminated by selecting only the median Curie temperature in order to retain an actual measured value of the $T_\mathrm{C}$. In DS1, we also added entries for each of the non-magnetic elements found in the data and set the Curie temperature to zero. After cleaning the raw data, DS1 contains the Curie temperatures of 2,557 unique compounds and DS2 contains 3,194. There is an overlap of 1,189 compounds between DS1 and DS2. Our feature vector has 85 features, each one describing a distinct element found in the data. Each compound is characterized by placing the percentage that each element occupies in the compound in the appropriate feature. For each machine-learning prediction, a randomly selected third of the compounds in DS1 was used as the test data and the other two thirds were used to train the models.

It is not possible to visualize the data in 85 dimensions. However, a t-distributed stochastic neighbor embedding\cite{tsne} (t-SNE) projection reduces the data to two dimensions and may reveal data clustering. Figure~\ref{fig:TCTSNE}(a) shows the t-SNE plot for DS1, where each point is \textcolor{black}{colored by the majority element in the compound.} \textcolor{black}{} The red circles show areas where compounds with the same majority element cluster together. Figure~\ref{fig:TCTSNE}(b) shows the same t-SNE plot but with the colors corresponding to the Curie temperatures of each compound. Each point in these plots represents a magnetic compound in the dataset. A comparison of the two plots reveal that most of the high-$T_\mathrm{C}$ materials have a majority element of either cobalt or iron. It also shows that there are occasional high-$T_\mathrm{C}$ spikes in clusters where the $T_\mathrm{C}$ is typically very low (for example, FeNi$_\mathrm{3}$ and Cr$_\mathrm{2}$Pt$_\mathrm{3}$). Figure~\ref{fig:TCTSNE-DS2} shows similar plots for DS2.

In Ref.~[\onlinecite{nelson2019predicting}], ridge regression, kernel ridge regression, neural networks, and random-forest machine-learning methods were all tested and the random-forest model made the most accurate predictions. We used DS1 in a $k$-nearest neighbors model and compared its performance to the accuracy of the random forest model. The $k$-NN algorithm was varied for 1 to 20 neighbors to determine the optimal $k$. A $k$-NN model with 2 neighbors provided best prediction accuracy but was not as good as the random forest (see Fig.~\ref{fig:NSdataPredictions}). 


To see how well the DS1 random forest model generalizes \textcolor{black}{ on \emph{unseen data}, we did a ``blind'' test} \textcolor{black}{} on all the data in DS2 (see Fig.~\ref{fig:Random Forest DS2}). 54\% of the predicted Curie temperatures were within 50\,K and 70\% within 100\,K. The mean absolute error was 90\,K. This \textcolor{black}{is a larger error than the CV score on DS1 data, however, the model still makes} \textcolor{black}{} relatively accurate predictions on around 2,000 previously unseen magnets. \textcolor{black}{We also used k-fold cross validation to show how the addition of DS2 data affects our predictive accuracy. 50 iterations of 3-fold cross validation shows that DS1 averages an MAE of 73 K and a standard deviation of 3.2. A combination of DS1 and DS2 averages an MAE of 71 K and a standard deviation of 2.3}

\begin{figure}
    \caption{Random forest prediction using features based on the composition and 20 of the features generated by the MAST-ML\cite{mastml} Python library. 62\% of the $T_\mathrm{C}$ predictions for DS1 were within 50 K of the actual values and 78\% were within 100 K The mean absolute error was 69 K. 57\% of the $T_\mathrm{C}$ predictions for the MAST-ML data were within 50 K of the actual values and 77\% were within 100 K The mean absolute error was 70 K.}
    \includegraphics[width=8.5cm]{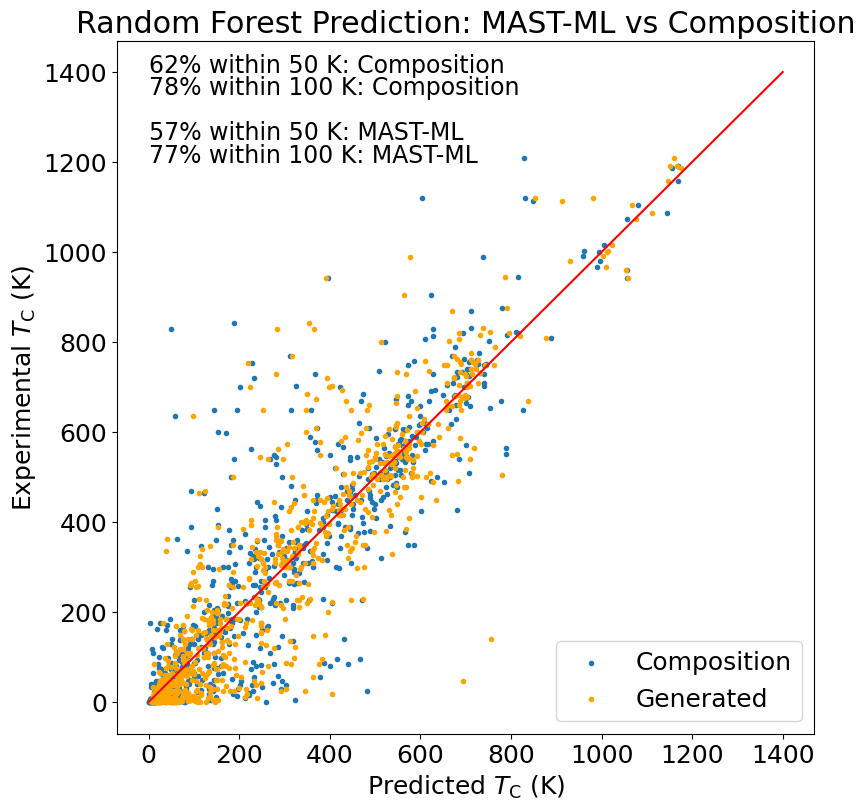}
    \label{fig:mastml}
\end{figure}

\begin{figure}

\caption{Random forest prediction with a balanced dataset. The Curie temperatures of the test data above 600 K were predicted with an accuracy of 37\% of the data within 50 K and 49\% within 100 K. The mean absolute error for those points was 141 K.}
\includegraphics[width=8cm]{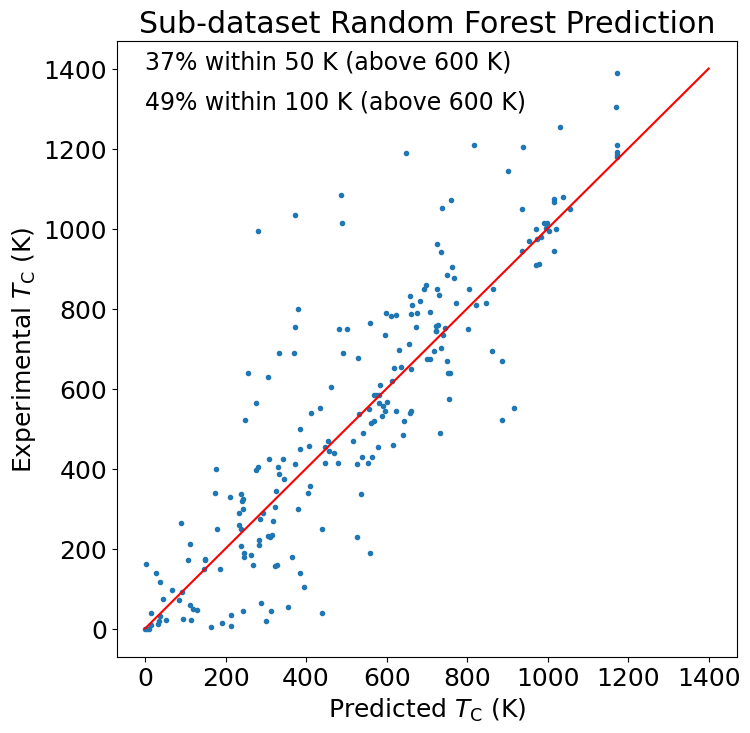}
\label{fig:subdataset}
\end{figure}

\begin{figure}
    \caption{MAE vs the amount of training data in a combined set of DS1 and DS2. For each iteration the test data was a new random sample of 850 compounds while the training data was randomly sampled from the remaining data.}
    \includegraphics[width=8.8cm]{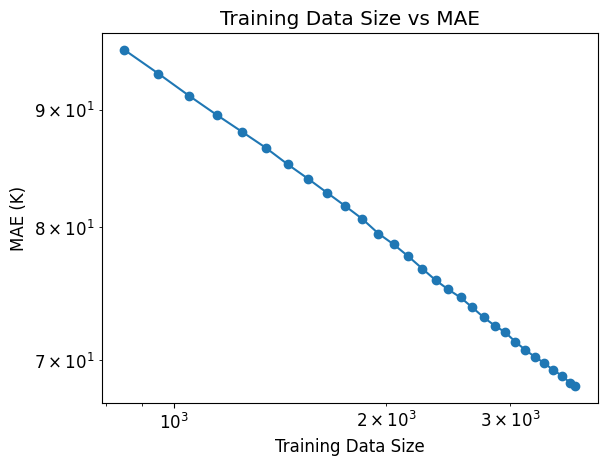}
    \label{fig:MAEvsTD}
\end{figure}

While the random-forest results are encouraging, a few different strategies were tried to improve the random-forest predictions. One of these strategies is to reduce the number of dimensions in the DS1 design matrix. A random forest was made for a range (5--85) of features using principal component analysis (PCA). PCA is not helpful to our model because, although noisy, the change in mean absolute error for each dataset shows that the accuracy improves up to about 60 features and stops improving after that as already observed in Ref.~[\onlinecite{nelson2019predicting}]. 

Another improvement strategy is designing better features. The MAST-ML\cite{mastml} Python library can generate descriptors based on about one hundred atomic properties (such as ionic radius, electronegativity, etc.). Using MAST-ML, we generated 428 features for the compounds in DS1 and selected the top 20 most meaningful ones identified through \textcolor{black}{MAST-ML's } \verb@EnsembleModelFeatureSelector@ \textcolor{black}{method which uses a random forest model to rank feature importance.} \textcolor{black}{} Figure~\ref{fig:mastml} shows these features used in a random forest and the comparison between these predictions and our random forest model predictions in Fig~\ref{fig:NSdataPredictions}. 
The MAST-ML features yield practically the same accuracy as with the original features. 
This is somewhat surprising because
the MAST-ML descriptors incorporate many atomic properties, not merely the composition. \textcolor{black}{However, reference~[\onlinecite{murdock2020domain}] shows that the domain knowledge in the generated features is not expected to improve prediction accuracy over simple fractional features when using large amounts of data. Our results suggest that DS1 is large enough to make an accurate model using our original fractional features.}  While the size of ML model using the MAST-ML descriptors is significantly smaller than the size of the model using the composition-only descriptors, both models have the same accuracy and both are efficient enough to be used in the large searches below. Our final models used the composition-only descriptors.
\begin{figure*}
    \floatbox[{\capbeside\thisfloatsetup{capbesideposition={right,center},capbesidewidth=4cm}}]{figure}[\FBwidth]
    {\caption{Eight plots showing the Curie temperatures for eight different binary compounds. Horizontal rows share the same scale on the y axis. The black lines show the random forest predicted Curie temperatures across the entire composition range (step size: +1 at - \%). The red dots are experimental data.\label{fig:TCsweep}}}
    {\includegraphics[width=12cm]{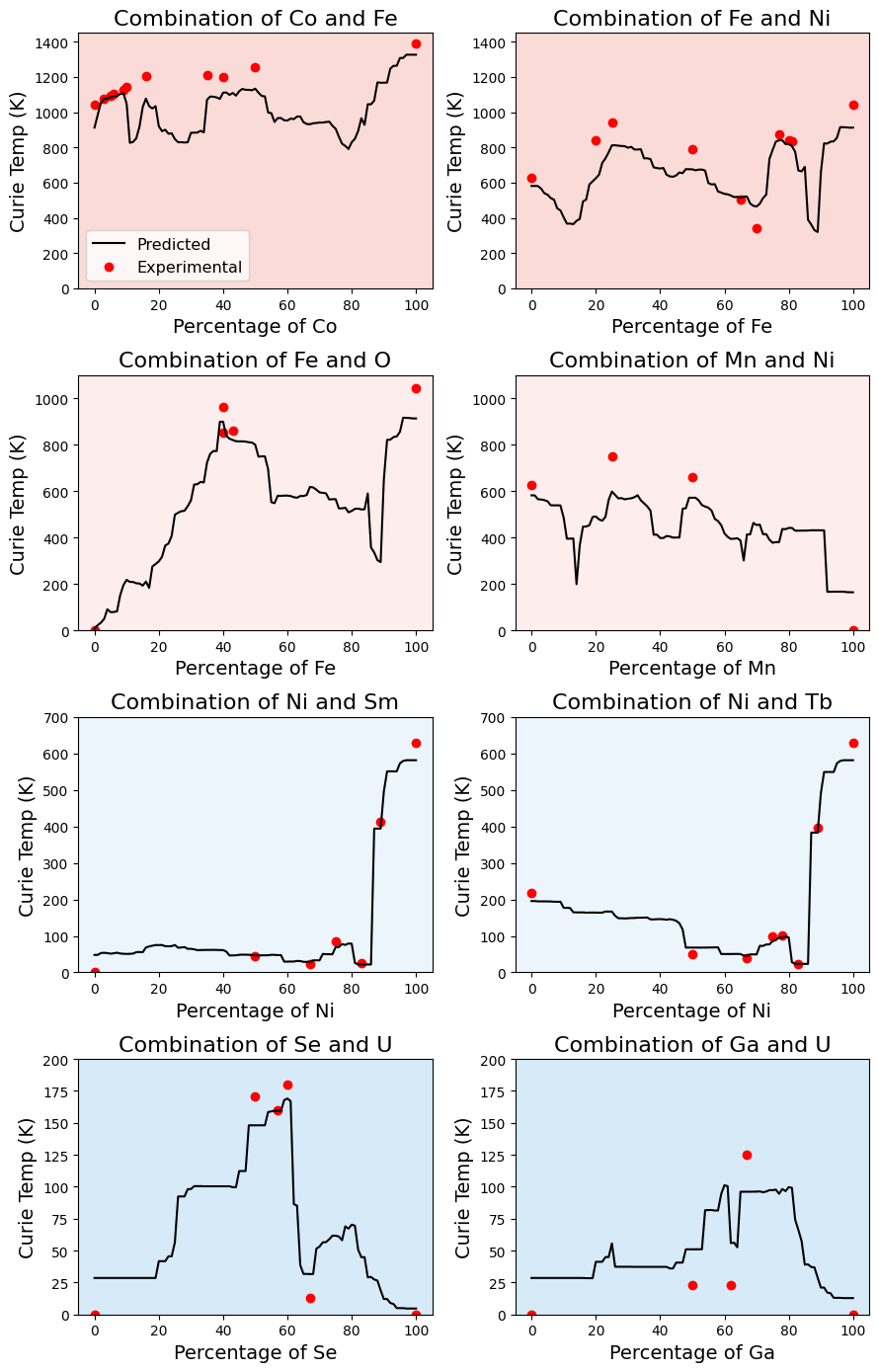}}
\end{figure*}

\begin{figure*}
    \begin{minipage}[b]{0.45\linewidth}
        \centering
        \includegraphics[width=\textwidth]{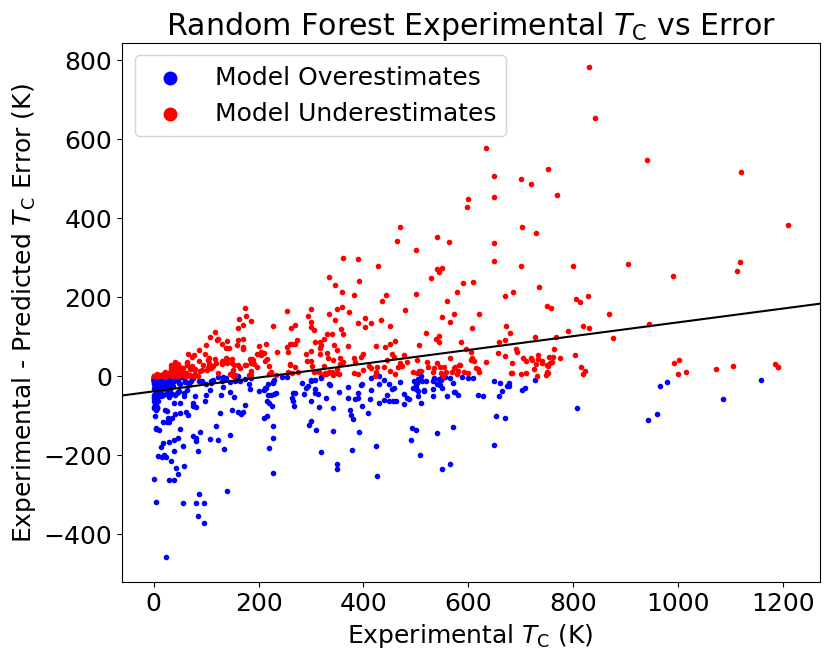}
        \label{fig:figure1}
    \end{minipage}
    \hspace{0.8cm}
    \begin{minipage}[b]{0.45\linewidth}
        \centering
        \includegraphics[width=\textwidth]{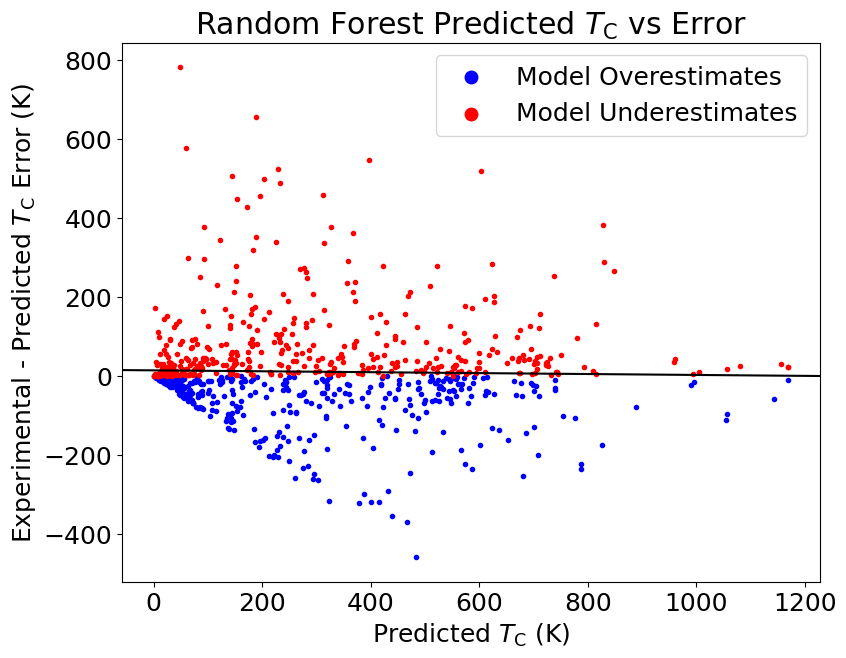}
        \label{fig:figure2}
    \end{minipage}
    \caption{Prediction error of the random forest model (experimental minus \emph{prediction}). (Left) Error versus \emph{experimental} $T_\mathrm{C}$ values. (Right) Error versus \emph{predicted} $T_\mathrm{C}$ values. The black line is the line of best fit. The errors are correlated with the experimental $T_\mathrm{C}$ but not with the predicted $T_\mathrm{C}$.}
    \label{fig:rf error}
\end{figure*}

Getting more balanced training data may be another way to improve the model. 
An analysis of DS1 shows a significant imbalance, 38\% of the magnetic materials have a $T_\mathrm{C}$ $0-100$ K, while only 15\% have a $T_\mathrm{C}$ $600-1400$ K. This could lead the model to reduce the prediction error at low temperatures at the expense of making larger errors at high temperatures. This is counter to the purpose of the model, which is that to discover candidate high-$T_\mathrm{C}$ magnets. 
Thus, we downsampled the low-$T_\mathrm{C}$ compounds to more evenly distribute DS1. The resulting model is shown in Fig.~\ref{fig:subdataset}. Even though there is a high density of low-$T_\mathrm{C}$ compounds, apparently they are necessary for accurate predictions of higher-$T_\mathrm{C}$ compounds. The balanced data predictions for high $T_\mathrm{C}$ materials are not as good as with the origin model, with an accuracy of 10\% less within 50\,K, 14\% less within 100\,K, and the mean absolute error 14\,K higher. 

Let us make one final comment before using the ML model to search for new high-$T_\mathrm{C}$. With this unprecedented large database of magnetic compounds, a large range of training set sizes are possible so that the learning rate of the model can be estimated. Figure~\ref{fig:MAEvsTD} shows the MAE on a test set of 850  compounds (approximately 1/3 the size of DS1, similar to the predictions in Fig.~\ref{fig:NSdataPredictions}) while the training set is increased, using the remaining data. The process was repeated 100 times and averaged. A linear fit reveals a slope of $\approx -0.22$. This learning rate is not particularly fast, but a better learning rate is probably not possible with composition-only descriptors. To reduce the MAE to 50\,K in the current model class would require a near tripling of the data.


The purpose of the ML model is to enable identification of candidate high-$T_\mathrm{C}$ materials. Our first attempt was to sweep over compositions of binary materials, but this revealed a systematic error. The error persists independent of the training data (DS1 or DS2), the particular ML method (random forest, $k$-NN), or the descriptors. 
Figure~\ref{fig:TCsweep} compares the predictions with the experimental $T_\mathrm{C}$ of eight different sets of binary systems. The top four red subplots analyze four different high-$T_\mathrm{C}$ materials and the four lower blue subplots analyze four different low-$T_\mathrm{C}$ materials. Each subplot shows the actual $T_\mathrm{C}$ value (red points) compared to the ML prediction (black line). A clear pattern emerges: between experimental data points, the ML model, instead of interpolating smoothly, drifts towards the average $T_\mathrm{C}$ of the training data (293 K). In the case of high-$T_\mathrm{C}$ materials, deep ``valleys'' occur in the absence of nearby experimental data. And for low-$T_\mathrm{C}$ materials, the model tends to overestimate between training points. In a pattern reminiscent of Gaussian processes, the model tends to ``regress toward the mean''.

This systematic drift toward the mean can be seen across all the test data in Fig.~\ref{fig:rf error}(left), which shows the difference between the predicted and the experimental $T_\mathrm{C}$ for the test data as a function of the experimental $T_\mathrm{C}$ (the random forest model using DS1). 
Intuition suggests that this systematic error could be mitigated simply by shifting and re-scaling according to the black line in the figure. However, the Fig.~\ref{fig:rf error}(right) demonstrates that this does not work. The errors and the \emph{predicted} $T_\mathrm{C}$'s are not correlated. The reader should remember that correlation is not transitive in general. Although the errors are correlated with the experimental $T_\mathrm{C}$'s, and the experimental and predicted $T_\mathrm{C}$'s are correlated, the errors and the \emph{predicted} $T_\mathrm{C}$'s are not necessarily correlated.

Recognizing the systematic error in predictions and being more interested in the accuracy of high-$T_\mathrm{C}$ predictions, we trained a random forest model on a combined set of DS1 and DS2 but only included the \textcolor{black}{967} compounds with $T_\mathrm{C}>600$~K thus boosting the mean of the training data from 293 K to 825 K and mitigating the regression to the mean. Obviously this new model cannot be used to predict low-$T_\mathrm{C}$ materials but as we are primarily concerned with high-$T_\mathrm{C}$ materials, this new model with a higher average protects the predictions from straying too far in data-deficient areas.

\begin{figure}
    \caption{Heat map showing the maximum predicted $T_\mathrm{C}$ of each combination of cobalt, iron, and $X_\mathrm{1}$ where $X_\mathrm{1}$ can be any element in materials with a $T_\mathrm{C}$ over 600 K found in DS1 and DS2 (excluding cobalt and iron). $T_\mathrm{C}$ is measured in Kelvins. Plot created using Python-Ternary.\cite{pythonternary}}
    \includegraphics[width=8cm]{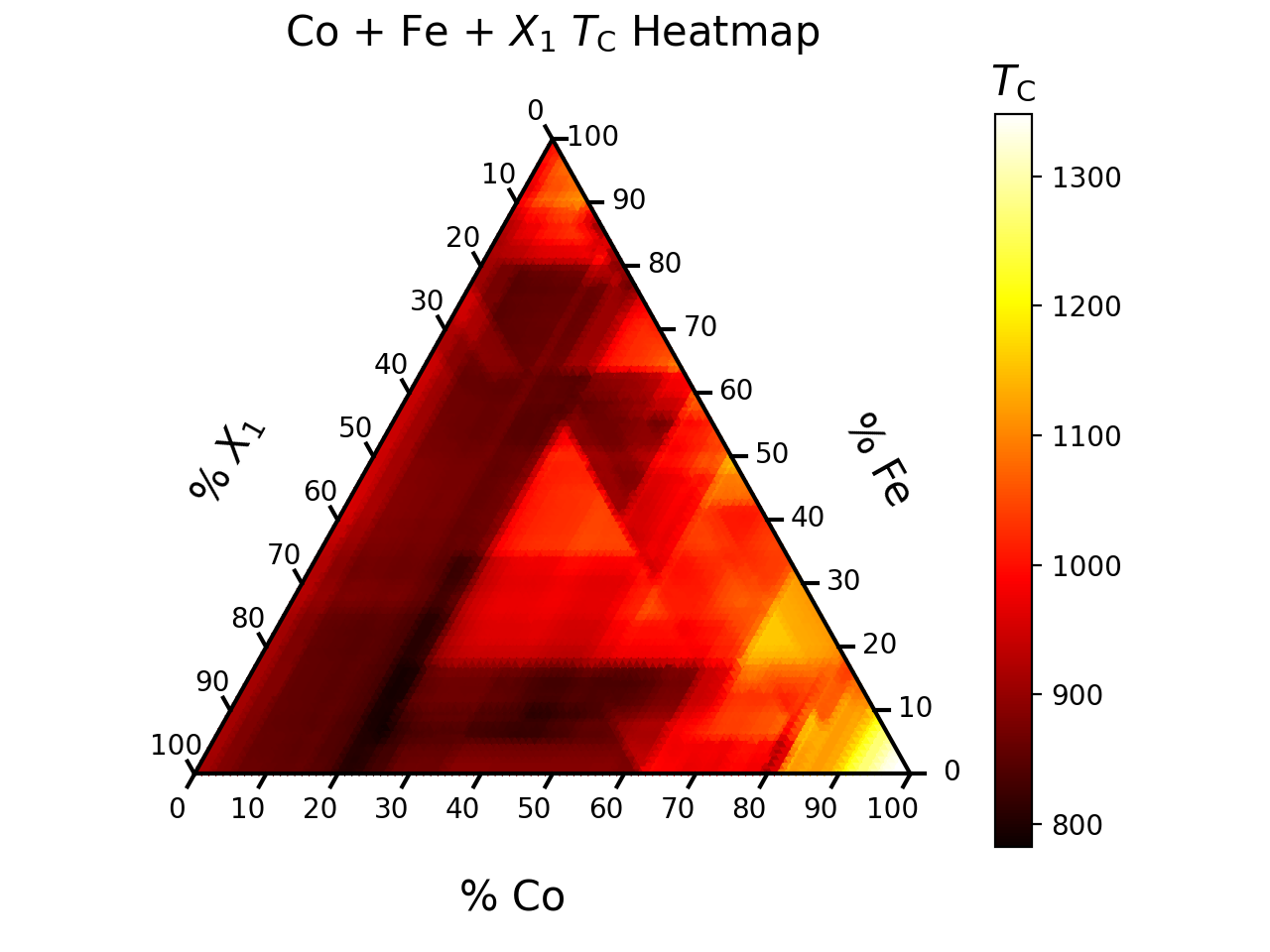}
    \label{fig:CoFeX ternary}
\end{figure}

\begin{figure}
    \caption{Heat map showing the maximum predicted $T_\mathrm{C}$ of each combination of iron, $X_\mathrm{1}$, and $X_\mathrm{2}$ where $X_\mathrm{1}$ and $X_\mathrm{2}$ can be any element in materials with a $T_\mathrm{C}$ over 600 K found in DS1 and DS2 excluding cobalt and iron. Each compound is composed of a minimum of 80\% iron. $T_\mathrm{C}$ is measured in Kelvins. Plot created using Python-Ternary.\cite{pythonternary} (This plot should not be directly compared with Fig. ~\ref{fig:CoFeX ternary} because of the difference in color scales.)}
    \includegraphics[width=8cm]{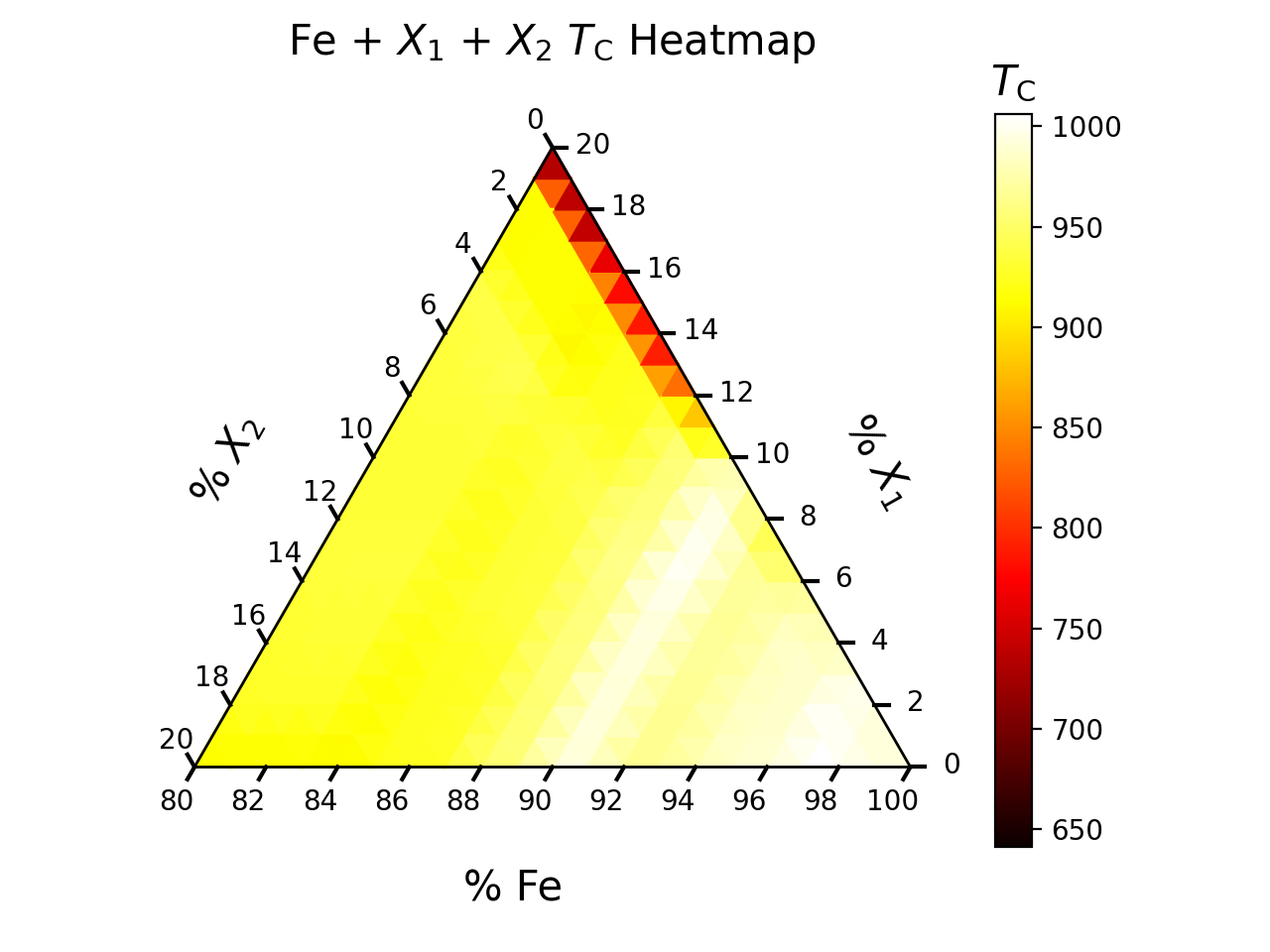}
    \label{fig:FE80XXX ternary}
\end{figure}

 This new high-$T_\mathrm{C}$-only model was used to search composition space for all binary and ternary candidate compounds with exceptionally high-$T_\mathrm{C}$. The high-$T_\mathrm{C}$-only model used only 71 features, restricted to elements that appear in compounds with $T_\mathrm{C}>600$~K. Figs~\ref{fig:TCTSNE} and~\ref{fig:TCTSNE-DS2} show that most of the highest-$T_\mathrm{C}$ compounds contain either cobalt, iron, or both. Using these elements, we generated every possible ternary combination containing iron, cobalt, and one of the other 69 elements ($X_\mathrm{1}$) in 1\% increments. The random forest model can rapidly predict $T_\mathrm{C}$ for this entire set. 
 Figure~\ref{fig:CoFeX ternary} displays the maximum predicted $T_\mathrm{C}$ of each combination of cobalt, iron, and $X_\mathrm{1}$. While cobalt clearly dominates the high-$T_\mathrm{C}$ compounds, materials with a high iron concentration also are predicted to have a high $T_\mathrm{C}$ and are worth exploring because iron is much cheaper than cobalt\cite{sciencebase}.  Figure~\ref{fig:FE80XXX ternary} shows similar predictions but for iron-rich ternary and binary candidates, which are inexpensive. 
 These predictions suggest that, at least in general, for maximizing $T_\mathrm{C}$, it is difficult to beat cobalt-rich materials, and that for minimizing materials cost for a relatively high $T_\mathrm{C}$, iron-rich materials are difficult to beat. Other searches excluding cobalt and iron were also performed but no competitive materials emerged.  

Unsuspected high-$T_\mathrm{C}$ materials sometimes appear. For example, see FeNi$_3$ and Cr$_2$Pt$_3$ in Fig.~\ref{fig:TCTSNE}---these turn up in clusters with very low Curie temperatures. But atomic structure, not merely composition, evidently plays an essential role in such materials. Devoid of explicit structure in the training data or the model, our predictions cannot be expected to include such candidates. Our models can learn a major portion of the relationship between chemical composition and Curie temperature and thus generally predict quite accurately. But it seems that finding exceptional cases will require data that includes structural information and a model that incorporates explicit relationships between magnetism and structure. As discussed at the beginning, the reliability of such models remains challenging, and corresponding data is limited, so computational discovery of new high-$T_\mathrm{C}$ remains an outstanding challenge.


\paragraph{Supplementary Material}
The supplementary material includes two error plots from the 2-NN predictions shown in the left plot of Fig.~\ref{fig:NSdataPredictions}. It also includes a plot of the number of features reduced by PCA vs. the MAE of the predictions made by a random forest model using that number of features.

\paragraph{Acknowledgements}
We would like to thank Mark Transtrum for useful discussions about the systematic error in Fig.~\ref{fig:rf error}. J.F.B. was generously supported by the College of Physical and Mathematical Sciences at Brigham Young University. J.F.B also acknowledges the mentorship provided by Ben Afflerbach and the Informatics Skunkworks Program. G.L.W.H. was supported by the U.S. National Science Foundation under Award \#DMR-1817321. V.T. acknowledges funding from the Critical Materials Institute, an Energy Innovation Hub funded by the U.S. Department of Energy Advanced Manufacturing Office. S.S. Acknowledges financial support from the Irish Research Council (IRCLA/2019/127).

\paragraph{Data Availability}
The data and code to reproduce the findings of this study are available here: \href{https://github.com/msg-byu/ML-for-CurieTemp-Predictions}{https://github.com/msg-byu/ML-for-CurieTemp-Predictions}.

\bibliography{rev1_final}

\providecommand{\noopsort}[1]{}\providecommand{\singleletter}[1]{#1}%
\begin{thebibliography}{42}%
\makeatletter
\providecommand \@ifxundefined [1]{%
 \@ifx{#1\undefined}
}%
\providecommand \@ifnum [1]{%
 \ifnum #1\expandafter \@firstoftwo
 \else \expandafter \@secondoftwo
 \fi
}%
\providecommand \@ifx [1]{%
 \ifx #1\expandafter \@firstoftwo
 \else \expandafter \@secondoftwo
 \fi
}%
\providecommand \natexlab [1]{#1}%
\providecommand \enquote  [1]{``#1''}%
\providecommand \bibnamefont  [1]{#1}%
\providecommand \bibfnamefont [1]{#1}%
\providecommand \citenamefont [1]{#1}%
\providecommand \href@noop [0]{\@secondoftwo}%
\providecommand \href [0]{\begingroup \@sanitize@url \@href}%
\providecommand \@href[1]{\@@startlink{#1}\@@href}%
\providecommand \@@href[1]{\endgroup#1\@@endlink}%
\providecommand \@sanitize@url [0]{\catcode `\\12\catcode `\$12\catcode
  `\&12\catcode `\#12\catcode `\^12\catcode `\_12\catcode `\%12\relax}%
\providecommand \@@startlink[1]{}%
\providecommand \@@endlink[0]{}%
\providecommand \url  [0]{\begingroup\@sanitize@url \@url }%
\providecommand \@url [1]{\endgroup\@href {#1}{\urlprefix }}%
\providecommand \urlprefix  [0]{URL }%
\providecommand \Eprint [0]{\href }%
\providecommand \doibase [0]{http://dx.doi.org/}%
\providecommand \selectlanguage [0]{\@gobble}%
\providecommand \bibinfo  [0]{\@secondoftwo}%
\providecommand \bibfield  [0]{\@secondoftwo}%
\providecommand \translation [1]{[#1]}%
\providecommand \BibitemOpen [0]{}%
\providecommand \bibitemStop [0]{}%
\providecommand \bibitemNoStop [0]{.\EOS\space}%
\providecommand \EOS [0]{\spacefactor3000\relax}%
\providecommand \BibitemShut  [1]{\csname bibitem#1\endcsname}%
\let\auto@bib@innerbib\@empty
\bibitem [{\citenamefont {Nelson}\ and\ \citenamefont
  {Sanvito}(2019)}]{nelson2019predicting}%
  \BibitemOpen
  \bibfield  {author} {\bibinfo {author} {\bibfnamefont {J.}~\bibnamefont
  {Nelson}}\ and\ \bibinfo {author} {\bibfnamefont {S.}~\bibnamefont
  {Sanvito}},\ }\bibfield  {title} {\enquote {\bibinfo {title} {Predicting the
  curie temperature of ferromagnets using machine learning},}\ }\href@noop {}
  {\bibfield  {journal} {\bibinfo  {journal} {Physical Review Materials}\
  }\textbf {\bibinfo {volume} {3}},\ \bibinfo {pages} {104405} (\bibinfo {year}
  {2019})}\BibitemShut {NoStop}%
\bibitem [{\citenamefont {Coey}(2020)}]{coey2020perspective}%
  \BibitemOpen
  \bibfield  {author} {\bibinfo {author} {\bibfnamefont {J.}~\bibnamefont
  {Coey}},\ }\bibfield  {title} {\enquote {\bibinfo {title} {Perspective and
  prospects for rare earth permanent magnets},}\ }\href@noop {} {\bibfield
  {journal} {\bibinfo  {journal} {Engineering}\ }\textbf {\bibinfo {volume}
  {6}},\ \bibinfo {pages} {119--131} (\bibinfo {year} {2020})}\BibitemShut
  {NoStop}%
\bibitem [{\citenamefont {Curtarolo}\ \emph {et~al.}(2013)\citenamefont
  {Curtarolo}, \citenamefont {Hart}, \citenamefont {Nardelli}, \citenamefont
  {Mingo}, \citenamefont {Sanvito},\ and\ \citenamefont
  {Levy}}]{curtarolo2013high}%
  \BibitemOpen
  \bibfield  {author} {\bibinfo {author} {\bibfnamefont {S.}~\bibnamefont
  {Curtarolo}}, \bibinfo {author} {\bibfnamefont {G.~L.}\ \bibnamefont {Hart}},
  \bibinfo {author} {\bibfnamefont {M.~B.}\ \bibnamefont {Nardelli}}, \bibinfo
  {author} {\bibfnamefont {N.}~\bibnamefont {Mingo}}, \bibinfo {author}
  {\bibfnamefont {S.}~\bibnamefont {Sanvito}}, \ and\ \bibinfo {author}
  {\bibfnamefont {O.}~\bibnamefont {Levy}},\ }\bibfield  {title} {\enquote
  {\bibinfo {title} {The high-throughput highway to computational materials
  design},}\ }\href@noop {} {\bibfield  {journal} {\bibinfo  {journal} {Nature
  materials}\ }\textbf {\bibinfo {volume} {12}},\ \bibinfo {pages} {191--201}
  (\bibinfo {year} {2013})}\BibitemShut {NoStop}%
\bibitem [{\citenamefont {Merker}\ \emph {et~al.}(2022)\citenamefont {Merker},
  \citenamefont {Heiberger}, \citenamefont {Nguyen}, \citenamefont {Liu},
  \citenamefont {Chen}, \citenamefont {Andrejevic}, \citenamefont {Drucker},
  \citenamefont {Okabe}, \citenamefont {Kim}, \citenamefont {Wang} \emph
  {et~al.}}]{merker2022machine}%
  \BibitemOpen
  \bibfield  {author} {\bibinfo {author} {\bibfnamefont {H.~A.}\ \bibnamefont
  {Merker}}, \bibinfo {author} {\bibfnamefont {H.}~\bibnamefont {Heiberger}},
  \bibinfo {author} {\bibfnamefont {L.}~\bibnamefont {Nguyen}}, \bibinfo
  {author} {\bibfnamefont {T.}~\bibnamefont {Liu}}, \bibinfo {author}
  {\bibfnamefont {Z.}~\bibnamefont {Chen}}, \bibinfo {author} {\bibfnamefont
  {N.}~\bibnamefont {Andrejevic}}, \bibinfo {author} {\bibfnamefont {N.~C.}\
  \bibnamefont {Drucker}}, \bibinfo {author} {\bibfnamefont {R.}~\bibnamefont
  {Okabe}}, \bibinfo {author} {\bibfnamefont {S.~E.}\ \bibnamefont {Kim}},
  \bibinfo {author} {\bibfnamefont {Y.}~\bibnamefont {Wang}},  \emph {et~al.},\
  }\bibfield  {title} {\enquote {\bibinfo {title} {Machine learning magnetism
  classifiers from atomic coordinates},}\ }\href@noop {} {\bibfield  {journal}
  {\bibinfo  {journal} {Iscience}\ }\textbf {\bibinfo {volume} {25}},\ \bibinfo
  {pages} {105192} (\bibinfo {year} {2022})}\BibitemShut {NoStop}%
\bibitem [{\citenamefont {Vedmedenko}\ \emph {et~al.}(2020)\citenamefont
  {Vedmedenko}, \citenamefont {Kawakami}, \citenamefont {Sheka}, \citenamefont
  {Gambardella}, \citenamefont {Kirilyuk}, \citenamefont {Hirohata},
  \citenamefont {Binek}, \citenamefont {Chubykalo-Fesenko}, \citenamefont
  {Sanvito}, \citenamefont {Kirby}, \citenamefont {Grollier}, \citenamefont
  {Everschor-Sitte}, \citenamefont {Kampfrath}, \citenamefont {You},\ and\
  \citenamefont {Berger}}]{Roadmap2020}%
  \BibitemOpen
  \bibfield  {author} {\bibinfo {author} {\bibfnamefont {E.~Y.}\ \bibnamefont
  {Vedmedenko}}, \bibinfo {author} {\bibfnamefont {R.~K.}\ \bibnamefont
  {Kawakami}}, \bibinfo {author} {\bibfnamefont {D.~D.}\ \bibnamefont {Sheka}},
  \bibinfo {author} {\bibfnamefont {P.}~\bibnamefont {Gambardella}}, \bibinfo
  {author} {\bibfnamefont {A.}~\bibnamefont {Kirilyuk}}, \bibinfo {author}
  {\bibfnamefont {A.}~\bibnamefont {Hirohata}}, \bibinfo {author}
  {\bibfnamefont {C.}~\bibnamefont {Binek}}, \bibinfo {author} {\bibfnamefont
  {O.}~\bibnamefont {Chubykalo-Fesenko}}, \bibinfo {author} {\bibfnamefont
  {S.}~\bibnamefont {Sanvito}}, \bibinfo {author} {\bibfnamefont {B.~J.}\
  \bibnamefont {Kirby}}, \bibinfo {author} {\bibfnamefont {J.}~\bibnamefont
  {Grollier}}, \bibinfo {author} {\bibfnamefont {K.}~\bibnamefont
  {Everschor-Sitte}}, \bibinfo {author} {\bibfnamefont {T.}~\bibnamefont
  {Kampfrath}}, \bibinfo {author} {\bibfnamefont {C.-Y.}\ \bibnamefont {You}},
  \ and\ \bibinfo {author} {\bibfnamefont {A.}~\bibnamefont {Berger}},\
  }\bibfield  {title} {\enquote {\bibinfo {title} {The 2020 magnetism
  roadmap},}\ }\href@noop {} {\bibfield  {journal} {\bibinfo  {journal} {J.
  Phys. D: Appl. Phys.}\ }\textbf {\bibinfo {volume} {53}},\ \bibinfo {pages}
  {453001} (\bibinfo {year} {2020})}\BibitemShut {NoStop}%
\bibitem [{\citenamefont {Coey}(2010)}]{coey2010magnetism}%
  \BibitemOpen
  \bibfield  {author} {\bibinfo {author} {\bibfnamefont {J.~M.}\ \bibnamefont
  {Coey}},\ }\href@noop {} {\emph {\bibinfo {title} {Magnetism and magnetic
  materials}}}\ (\bibinfo  {publisher} {Cambridge university press},\ \bibinfo
  {year} {2010})\BibitemShut {NoStop}%
\bibitem [{\citenamefont {Byland}\ \emph {et~al.}(2022)\citenamefont {Byland},
  \citenamefont {Shi}, \citenamefont {Parker}, \citenamefont {Zhao},
  \citenamefont {Ding}, \citenamefont {Mata}, \citenamefont {Magliari},
  \citenamefont {Palasyuk}, \citenamefont {Bud'ko}, \citenamefont {Canfield}
  \emph {et~al.}}]{byland2022statistics}%
  \BibitemOpen
  \bibfield  {author} {\bibinfo {author} {\bibfnamefont {J.~K.}\ \bibnamefont
  {Byland}}, \bibinfo {author} {\bibfnamefont {Y.}~\bibnamefont {Shi}},
  \bibinfo {author} {\bibfnamefont {D.~S.}\ \bibnamefont {Parker}}, \bibinfo
  {author} {\bibfnamefont {J.}~\bibnamefont {Zhao}}, \bibinfo {author}
  {\bibfnamefont {S.}~\bibnamefont {Ding}}, \bibinfo {author} {\bibfnamefont
  {R.}~\bibnamefont {Mata}}, \bibinfo {author} {\bibfnamefont {H.~E.}\
  \bibnamefont {Magliari}}, \bibinfo {author} {\bibfnamefont {A.}~\bibnamefont
  {Palasyuk}}, \bibinfo {author} {\bibfnamefont {S.~L.}\ \bibnamefont
  {Bud'ko}}, \bibinfo {author} {\bibfnamefont {P.~C.}\ \bibnamefont
  {Canfield}},  \emph {et~al.},\ }\bibfield  {title} {\enquote {\bibinfo
  {title} {Statistics on magnetic properties of co compounds: A database-driven
  method for discovering co-based ferromagnets},}\ }\href@noop {} {\bibfield
  {journal} {\bibinfo  {journal} {Physical Review Materials}\ }\textbf
  {\bibinfo {volume} {6}},\ \bibinfo {pages} {063803} (\bibinfo {year}
  {2022})}\BibitemShut {NoStop}%
\bibitem [{\citenamefont {Kabiraj}, \citenamefont {Kumar},\ and\ \citenamefont
  {Mahapatra}(2020)}]{kabiraj2020high}%
  \BibitemOpen
  \bibfield  {author} {\bibinfo {author} {\bibfnamefont {A.}~\bibnamefont
  {Kabiraj}}, \bibinfo {author} {\bibfnamefont {M.}~\bibnamefont {Kumar}}, \
  and\ \bibinfo {author} {\bibfnamefont {S.}~\bibnamefont {Mahapatra}},\
  }\bibfield  {title} {\enquote {\bibinfo {title} {High-throughput discovery of
  high curie point two-dimensional ferromagnetic materials},}\ }\href@noop {}
  {\bibfield  {journal} {\bibinfo  {journal} {npj Computational Materials}\
  }\textbf {\bibinfo {volume} {6}},\ \bibinfo {pages} {35} (\bibinfo {year}
  {2020})}\BibitemShut {NoStop}%
\bibitem [{\citenamefont {Graf}, \citenamefont {Felser},\ and\ \citenamefont
  {Parkin}(2011)}]{graf2011simple}%
  \BibitemOpen
  \bibfield  {author} {\bibinfo {author} {\bibfnamefont {T.}~\bibnamefont
  {Graf}}, \bibinfo {author} {\bibfnamefont {C.}~\bibnamefont {Felser}}, \ and\
  \bibinfo {author} {\bibfnamefont {S.~S.}\ \bibnamefont {Parkin}},\ }\bibfield
   {title} {\enquote {\bibinfo {title} {Simple rules for the understanding of
  heusler compounds},}\ }\href@noop {} {\bibfield  {journal} {\bibinfo
  {journal} {Progress in solid state chemistry}\ }\textbf {\bibinfo {volume}
  {39}},\ \bibinfo {pages} {1--50} (\bibinfo {year} {2011})}\BibitemShut
  {NoStop}%
\bibitem [{\citenamefont {Sanvito}\ \emph
  {et~al.}(2017{\natexlab{a}})\citenamefont {Sanvito}, \citenamefont {Oses},
  \citenamefont {Xue}, \citenamefont {Tiwari}, \citenamefont {Zic},
  \citenamefont {Archer}, \citenamefont {Tozman}, \citenamefont {Venkatesan},
  \citenamefont {Coey},\ and\ \citenamefont
  {Curtarolo}}]{doi:10.1126/sciadv.1602241}%
  \BibitemOpen
  \bibfield  {author} {\bibinfo {author} {\bibfnamefont {S.}~\bibnamefont
  {Sanvito}}, \bibinfo {author} {\bibfnamefont {C.}~\bibnamefont {Oses}},
  \bibinfo {author} {\bibfnamefont {J.}~\bibnamefont {Xue}}, \bibinfo {author}
  {\bibfnamefont {A.}~\bibnamefont {Tiwari}}, \bibinfo {author} {\bibfnamefont
  {M.}~\bibnamefont {Zic}}, \bibinfo {author} {\bibfnamefont {T.}~\bibnamefont
  {Archer}}, \bibinfo {author} {\bibfnamefont {P.}~\bibnamefont {Tozman}},
  \bibinfo {author} {\bibfnamefont {M.}~\bibnamefont {Venkatesan}}, \bibinfo
  {author} {\bibfnamefont {M.}~\bibnamefont {Coey}}, \ and\ \bibinfo {author}
  {\bibfnamefont {S.}~\bibnamefont {Curtarolo}},\ }\bibfield  {title} {\enquote
  {\bibinfo {title} {Accelerated discovery of new magnets in the heusler alloy
  family},}\ }\href {\doibase 10.1126/sciadv.1602241} {\bibfield  {journal}
  {\bibinfo  {journal} {Science Advances}\ }\textbf {\bibinfo {volume} {3}},\
  \bibinfo {pages} {e1602241} (\bibinfo {year} {2017}{\natexlab{a}})},\ \Eprint
  {http://arxiv.org/abs/https://www.science.org/doi/pdf/10.1126/sciadv.1602241}
  {https://www.science.org/doi/pdf/10.1126/sciadv.1602241} \BibitemShut
  {NoStop}%
\bibitem [{\citenamefont {Nolting}, \citenamefont {Vega},\ and\ \citenamefont
  {Fauster}(1995)}]{nolting1995electronic}%
  \BibitemOpen
  \bibfield  {author} {\bibinfo {author} {\bibfnamefont {W.}~\bibnamefont
  {Nolting}}, \bibinfo {author} {\bibfnamefont {A.}~\bibnamefont {Vega}}, \
  and\ \bibinfo {author} {\bibfnamefont {T.}~\bibnamefont {Fauster}},\
  }\bibfield  {title} {\enquote {\bibinfo {title} {Electronic quasiparticle
  structure of ferromagnetic bcc iron},}\ }\href@noop {} {\bibfield  {journal}
  {\bibinfo  {journal} {Zeitschrift f{\"u}r Physik B Condensed Matter}\
  }\textbf {\bibinfo {volume} {96}},\ \bibinfo {pages} {357--372} (\bibinfo
  {year} {1995})}\BibitemShut {NoStop}%
\bibitem [{\citenamefont {K{\"u}bler}(2006)}]{kubler2006ab}%
  \BibitemOpen
  \bibfield  {author} {\bibinfo {author} {\bibfnamefont {J.}~\bibnamefont
  {K{\"u}bler}},\ }\bibfield  {title} {\enquote {\bibinfo {title} {Ab initio
  estimates of the curie temperature for magnetic compounds},}\ }\href@noop {}
  {\bibfield  {journal} {\bibinfo  {journal} {Journal of Physics: Condensed
  Matter}\ }\textbf {\bibinfo {volume} {18}},\ \bibinfo {pages} {9795}
  (\bibinfo {year} {2006})}\BibitemShut {NoStop}%
\bibitem [{\citenamefont {K{\"u}bler}, \citenamefont {Fecher},\ and\
  \citenamefont {Felser}(2007)}]{kubler2007understanding}%
  \BibitemOpen
  \bibfield  {author} {\bibinfo {author} {\bibfnamefont {J.}~\bibnamefont
  {K{\"u}bler}}, \bibinfo {author} {\bibfnamefont {G.}~\bibnamefont {Fecher}},
  \ and\ \bibinfo {author} {\bibfnamefont {C.}~\bibnamefont {Felser}},\
  }\bibfield  {title} {\enquote {\bibinfo {title} {Understanding the trend in
  the curie temperatures of co 2-based heusler compounds: Ab initio
  calculations},}\ }\href@noop {} {\bibfield  {journal} {\bibinfo  {journal}
  {Physical Review B}\ }\textbf {\bibinfo {volume} {76}},\ \bibinfo {pages}
  {024414} (\bibinfo {year} {2007})}\BibitemShut {NoStop}%
\bibitem [{\citenamefont {Halilov}\ \emph {et~al.}(1997)\citenamefont
  {Halilov}, \citenamefont {Perlov}, \citenamefont {Oppeneer},\ and\
  \citenamefont {Eschrig}}]{halilov1997magnon}%
  \BibitemOpen
  \bibfield  {author} {\bibinfo {author} {\bibfnamefont {S.}~\bibnamefont
  {Halilov}}, \bibinfo {author} {\bibfnamefont {A.}~\bibnamefont {Perlov}},
  \bibinfo {author} {\bibfnamefont {P.}~\bibnamefont {Oppeneer}}, \ and\
  \bibinfo {author} {\bibfnamefont {H.}~\bibnamefont {Eschrig}},\ }\bibfield
  {title} {\enquote {\bibinfo {title} {Magnon spectrum and related
  finite-temperature magnetic properties: A first-principle approach},}\
  }\href@noop {} {\bibfield  {journal} {\bibinfo  {journal} {Europhysics
  Letters}\ }\textbf {\bibinfo {volume} {39}},\ \bibinfo {pages} {91} (\bibinfo
  {year} {1997})}\BibitemShut {NoStop}%
\bibitem [{\citenamefont {K{\"o}rmann}\ \emph {et~al.}(2008)\citenamefont
  {K{\"o}rmann}, \citenamefont {Dick}, \citenamefont {Grabowski}, \citenamefont
  {Hallstedt}, \citenamefont {Hickel},\ and\ \citenamefont
  {Neugebauer}}]{kormann2008free}%
  \BibitemOpen
  \bibfield  {author} {\bibinfo {author} {\bibfnamefont {F.}~\bibnamefont
  {K{\"o}rmann}}, \bibinfo {author} {\bibfnamefont {A.}~\bibnamefont {Dick}},
  \bibinfo {author} {\bibfnamefont {B.}~\bibnamefont {Grabowski}}, \bibinfo
  {author} {\bibfnamefont {B.}~\bibnamefont {Hallstedt}}, \bibinfo {author}
  {\bibfnamefont {T.}~\bibnamefont {Hickel}}, \ and\ \bibinfo {author}
  {\bibfnamefont {J.}~\bibnamefont {Neugebauer}},\ }\bibfield  {title}
  {\enquote {\bibinfo {title} {Free energy of bcc iron: Integrated ab initio
  derivation of vibrational, electronic, and magnetic contributions},}\
  }\href@noop {} {\bibfield  {journal} {\bibinfo  {journal} {Physical Review
  B}\ }\textbf {\bibinfo {volume} {78}},\ \bibinfo {pages} {033102} (\bibinfo
  {year} {2008})}\BibitemShut {NoStop}%
\bibitem [{\citenamefont {Rosengaard}\ and\ \citenamefont
  {Johansson}(1997)}]{PhysRevB.55.14975}%
  \BibitemOpen
  \bibfield  {author} {\bibinfo {author} {\bibfnamefont {N.~M.}\ \bibnamefont
  {Rosengaard}}\ and\ \bibinfo {author} {\bibfnamefont {B.}~\bibnamefont
  {Johansson}},\ }\bibfield  {title} {\enquote {\bibinfo {title}
  {Finite-temperature study of itinerant ferromagnetism in fe, co, and ni},}\
  }\href {\doibase 10.1103/PhysRevB.55.14975} {\bibfield  {journal} {\bibinfo
  {journal} {Phys. Rev. B}\ }\textbf {\bibinfo {volume} {55}},\ \bibinfo
  {pages} {14975--14986} (\bibinfo {year} {1997})}\BibitemShut {NoStop}%
\bibitem [{\citenamefont {Le{\v{z}}ai{\'c}}, \citenamefont {Mavropoulos},\ and\
  \citenamefont {Bl{\"u}gel}(2007)}]{levzaic2007first}%
  \BibitemOpen
  \bibfield  {author} {\bibinfo {author} {\bibfnamefont {M.}~\bibnamefont
  {Le{\v{z}}ai{\'c}}}, \bibinfo {author} {\bibfnamefont {P.}~\bibnamefont
  {Mavropoulos}}, \ and\ \bibinfo {author} {\bibfnamefont {S.}~\bibnamefont
  {Bl{\"u}gel}},\ }\bibfield  {title} {\enquote {\bibinfo {title}
  {First-principles prediction of high curie temperature for ferromagnetic
  bcc-co and bcc-feco alloys and its relevance to tunneling
  magnetoresistance},}\ }\href@noop {} {\bibfield  {journal} {\bibinfo
  {journal} {Applied physics letters}\ }\textbf {\bibinfo {volume} {90}},\
  \bibinfo {pages} {082504} (\bibinfo {year} {2007})}\BibitemShut {NoStop}%
\bibitem [{\citenamefont {Rusz}\ \emph {et~al.}(2006)\citenamefont {Rusz},
  \citenamefont {Bergqvist}, \citenamefont {Kudrnovsk\'y},\ and\ \citenamefont
  {Turek}}]{PhysRevB.73.214412}%
  \BibitemOpen
  \bibfield  {author} {\bibinfo {author} {\bibfnamefont {J.}~\bibnamefont
  {Rusz}}, \bibinfo {author} {\bibfnamefont {L.}~\bibnamefont {Bergqvist}},
  \bibinfo {author} {\bibfnamefont {J.}~\bibnamefont {Kudrnovsk\'y}}, \ and\
  \bibinfo {author} {\bibfnamefont {I.}~\bibnamefont {Turek}},\ }\bibfield
  {title} {\enquote {\bibinfo {title} {Exchange interactions and curie
  temperatures in ${\mathrm{ni}}_{2\ensuremath{-}x}\mathrm{Mn}\mathrm{Sb}$
  alloys: First-principles study},}\ }\href {\doibase
  10.1103/PhysRevB.73.214412} {\bibfield  {journal} {\bibinfo  {journal} {Phys.
  Rev. B}\ }\textbf {\bibinfo {volume} {73}},\ \bibinfo {pages} {214412}
  (\bibinfo {year} {2006})}\BibitemShut {NoStop}%
\bibitem [{\citenamefont {Ruban}\ \emph {et~al.}(2007)\citenamefont {Ruban},
  \citenamefont {Khmelevskyi}, \citenamefont {Mohn},\ and\ \citenamefont
  {Johansson}}]{PhysRevB.75.054402}%
  \BibitemOpen
  \bibfield  {author} {\bibinfo {author} {\bibfnamefont {A.~V.}\ \bibnamefont
  {Ruban}}, \bibinfo {author} {\bibfnamefont {S.}~\bibnamefont {Khmelevskyi}},
  \bibinfo {author} {\bibfnamefont {P.}~\bibnamefont {Mohn}}, \ and\ \bibinfo
  {author} {\bibfnamefont {B.}~\bibnamefont {Johansson}},\ }\bibfield  {title}
  {\enquote {\bibinfo {title} {Temperature-induced longitudinal spin
  fluctuations in fe and ni},}\ }\href {\doibase 10.1103/PhysRevB.75.054402}
  {\bibfield  {journal} {\bibinfo  {journal} {Phys. Rev. B}\ }\textbf {\bibinfo
  {volume} {75}},\ \bibinfo {pages} {054402} (\bibinfo {year}
  {2007})}\BibitemShut {NoStop}%
\bibitem [{\citenamefont {Tanaka}\ and\ \citenamefont
  {Gohda}(2020)}]{tanaka2020prediction}%
  \BibitemOpen
  \bibfield  {author} {\bibinfo {author} {\bibfnamefont {T.}~\bibnamefont
  {Tanaka}}\ and\ \bibinfo {author} {\bibfnamefont {Y.}~\bibnamefont {Gohda}},\
  }\bibfield  {title} {\enquote {\bibinfo {title} {Prediction of the curie
  temperature considering the dependence of the phonon free energy on magnetic
  states},}\ }\href@noop {} {\bibfield  {journal} {\bibinfo  {journal} {npj
  Computational Materials}\ }\textbf {\bibinfo {volume} {6}},\ \bibinfo {pages}
  {184} (\bibinfo {year} {2020})}\BibitemShut {NoStop}%
\bibitem [{\citenamefont {K{\"o}rmann}\ \emph {et~al.}(2009)\citenamefont
  {K{\"o}rmann}, \citenamefont {Dick}, \citenamefont {Hickel},\ and\
  \citenamefont {Neugebauer}}]{kormann2009pressure}%
  \BibitemOpen
  \bibfield  {author} {\bibinfo {author} {\bibfnamefont {F.}~\bibnamefont
  {K{\"o}rmann}}, \bibinfo {author} {\bibfnamefont {A.}~\bibnamefont {Dick}},
  \bibinfo {author} {\bibfnamefont {T.}~\bibnamefont {Hickel}}, \ and\ \bibinfo
  {author} {\bibfnamefont {J.}~\bibnamefont {Neugebauer}},\ }\bibfield  {title}
  {\enquote {\bibinfo {title} {Pressure dependence of the curie temperature in
  bcc iron studied by ab initio simulations},}\ }\href@noop {} {\bibfield
  {journal} {\bibinfo  {journal} {Physical Review B}\ }\textbf {\bibinfo
  {volume} {79}},\ \bibinfo {pages} {184406} (\bibinfo {year}
  {2009})}\BibitemShut {NoStop}%
\bibitem [{\citenamefont {Liechtenstein}\ \emph {et~al.}(1987)\citenamefont
  {Liechtenstein}, \citenamefont {Katsnelson}, \citenamefont {Antropov},\ and\
  \citenamefont {Gubanov}}]{liechtenstein1987local}%
  \BibitemOpen
  \bibfield  {author} {\bibinfo {author} {\bibfnamefont {A.~I.}\ \bibnamefont
  {Liechtenstein}}, \bibinfo {author} {\bibfnamefont {M.}~\bibnamefont
  {Katsnelson}}, \bibinfo {author} {\bibfnamefont {V.}~\bibnamefont
  {Antropov}}, \ and\ \bibinfo {author} {\bibfnamefont {V.}~\bibnamefont
  {Gubanov}},\ }\bibfield  {title} {\enquote {\bibinfo {title} {Local spin
  density functional approach to the theory of exchange interactions in
  ferromagnetic metals and alloys},}\ }\href@noop {} {\bibfield  {journal}
  {\bibinfo  {journal} {Journal of Magnetism and Magnetic Materials}\ }\textbf
  {\bibinfo {volume} {67}},\ \bibinfo {pages} {65--74} (\bibinfo {year}
  {1987})}\BibitemShut {NoStop}%
\bibitem [{\citenamefont {Pajda}\ \emph {et~al.}(2001)\citenamefont {Pajda},
  \citenamefont {Kudrnovsk{\`y}}, \citenamefont {Turek}, \citenamefont
  {Drchal},\ and\ \citenamefont {Bruno}}]{pajda2001ab}%
  \BibitemOpen
  \bibfield  {author} {\bibinfo {author} {\bibfnamefont {M.}~\bibnamefont
  {Pajda}}, \bibinfo {author} {\bibfnamefont {J.}~\bibnamefont
  {Kudrnovsk{\`y}}}, \bibinfo {author} {\bibfnamefont {I.}~\bibnamefont
  {Turek}}, \bibinfo {author} {\bibfnamefont {V.}~\bibnamefont {Drchal}}, \
  and\ \bibinfo {author} {\bibfnamefont {P.}~\bibnamefont {Bruno}},\ }\bibfield
   {title} {\enquote {\bibinfo {title} {Ab initio calculations of exchange
  interactions, spin-wave stiffness constants, and curie temperatures of fe,
  co, and ni},}\ }\href@noop {} {\bibfield  {journal} {\bibinfo  {journal}
  {Physical Review B}\ }\textbf {\bibinfo {volume} {64}},\ \bibinfo {pages}
  {174402} (\bibinfo {year} {2001})}\BibitemShut {NoStop}%
\bibitem [{\citenamefont {Turek}, \citenamefont {Rusz},\ and\ \citenamefont
  {Divi{\v{s}}}(2005)}]{turek2005electronic}%
  \BibitemOpen
  \bibfield  {author} {\bibinfo {author} {\bibfnamefont {I.}~\bibnamefont
  {Turek}}, \bibinfo {author} {\bibfnamefont {J.}~\bibnamefont {Rusz}}, \ and\
  \bibinfo {author} {\bibfnamefont {M.}~\bibnamefont {Divi{\v{s}}}},\
  }\bibfield  {title} {\enquote {\bibinfo {title} {Electronic structure and
  volume magnetostriction of rare-earth metals and compounds},}\ }\href@noop {}
  {\bibfield  {journal} {\bibinfo  {journal} {Journal of magnetism and magnetic
  materials}\ }\textbf {\bibinfo {volume} {290}},\ \bibinfo {pages} {357--363}
  (\bibinfo {year} {2005})}\BibitemShut {NoStop}%
\bibitem [{\citenamefont {Gong}\ \emph {et~al.}(2019)\citenamefont {Gong},
  \citenamefont {Yi}, \citenamefont {Evans}, \citenamefont {Xu},\ and\
  \citenamefont {Gutfleisch}}]{gong2019calculating}%
  \BibitemOpen
  \bibfield  {author} {\bibinfo {author} {\bibfnamefont {Q.}~\bibnamefont
  {Gong}}, \bibinfo {author} {\bibfnamefont {M.}~\bibnamefont {Yi}}, \bibinfo
  {author} {\bibfnamefont {R.~F.}\ \bibnamefont {Evans}}, \bibinfo {author}
  {\bibfnamefont {B.-X.}\ \bibnamefont {Xu}}, \ and\ \bibinfo {author}
  {\bibfnamefont {O.}~\bibnamefont {Gutfleisch}},\ }\bibfield  {title}
  {\enquote {\bibinfo {title} {Calculating temperature-dependent properties of
  nd 2 fe 14 b permanent magnets by atomistic spin model simulations},}\
  }\href@noop {} {\bibfield  {journal} {\bibinfo  {journal} {Physical Review
  B}\ }\textbf {\bibinfo {volume} {99}},\ \bibinfo {pages} {214409} (\bibinfo
  {year} {2019})}\BibitemShut {NoStop}%
\bibitem [{\citenamefont {Takahashi}, \citenamefont {Ogura},\ and\
  \citenamefont {Akai}(2007)}]{takahashi2007first}%
  \BibitemOpen
  \bibfield  {author} {\bibinfo {author} {\bibfnamefont {C.}~\bibnamefont
  {Takahashi}}, \bibinfo {author} {\bibfnamefont {M.}~\bibnamefont {Ogura}}, \
  and\ \bibinfo {author} {\bibfnamefont {H.}~\bibnamefont {Akai}},\ }\bibfield
  {title} {\enquote {\bibinfo {title} {First-principles calculation of the
  curie temperature slater--pauling curve},}\ }\href@noop {} {\bibfield
  {journal} {\bibinfo  {journal} {Journal of Physics: Condensed Matter}\
  }\textbf {\bibinfo {volume} {19}},\ \bibinfo {pages} {365233} (\bibinfo
  {year} {2007})}\BibitemShut {NoStop}%
\bibitem [{\citenamefont {Song}\ \emph {et~al.}(2020)\citenamefont {Song},
  \citenamefont {Chen}, \citenamefont {Meng}, \citenamefont {Cheng},
  \citenamefont {Wang}, \citenamefont {Sun},\ and\ \citenamefont
  {Yin}}]{song2020machine}%
  \BibitemOpen
  \bibfield  {author} {\bibinfo {author} {\bibfnamefont {Z.}~\bibnamefont
  {Song}}, \bibinfo {author} {\bibfnamefont {X.}~\bibnamefont {Chen}}, \bibinfo
  {author} {\bibfnamefont {F.}~\bibnamefont {Meng}}, \bibinfo {author}
  {\bibfnamefont {G.}~\bibnamefont {Cheng}}, \bibinfo {author} {\bibfnamefont
  {C.}~\bibnamefont {Wang}}, \bibinfo {author} {\bibfnamefont {Z.}~\bibnamefont
  {Sun}}, \ and\ \bibinfo {author} {\bibfnamefont {W.-J.}\ \bibnamefont
  {Yin}},\ }\bibfield  {title} {\enquote {\bibinfo {title} {Machine learning in
  materials design: Algorithm and application},}\ }\href@noop {} {\bibfield
  {journal} {\bibinfo  {journal} {Chinese Physics B}\ }\textbf {\bibinfo
  {volume} {29}},\ \bibinfo {pages} {116103} (\bibinfo {year}
  {2020})}\BibitemShut {NoStop}%
\bibitem [{\citenamefont {Sanvito}\ \emph
  {et~al.}(2017{\natexlab{b}})\citenamefont {Sanvito}, \citenamefont {Oses},
  \citenamefont {Xue}, \citenamefont {Tiwari}, \citenamefont {Zic},
  \citenamefont {Archer}, \citenamefont {Tozman}, \citenamefont {Venkatesan},
  \citenamefont {Coey},\ and\ \citenamefont {Curtarolo}}]{Sanvito2017SA}%
  \BibitemOpen
  \bibfield  {author} {\bibinfo {author} {\bibfnamefont {S.}~\bibnamefont
  {Sanvito}}, \bibinfo {author} {\bibfnamefont {C.}~\bibnamefont {Oses}},
  \bibinfo {author} {\bibfnamefont {J.}~\bibnamefont {Xue}}, \bibinfo {author}
  {\bibfnamefont {A.}~\bibnamefont {Tiwari}}, \bibinfo {author} {\bibfnamefont
  {M.}~\bibnamefont {Zic}}, \bibinfo {author} {\bibfnamefont {T.}~\bibnamefont
  {Archer}}, \bibinfo {author} {\bibfnamefont {P.}~\bibnamefont {Tozman}},
  \bibinfo {author} {\bibfnamefont {M.}~\bibnamefont {Venkatesan}}, \bibinfo
  {author} {\bibfnamefont {M.}~\bibnamefont {Coey}}, \ and\ \bibinfo {author}
  {\bibfnamefont {S.}~\bibnamefont {Curtarolo}},\ }\bibfield  {title} {\enquote
  {\bibinfo {title} {Accelerated discovery of new magnets in the heusler alloy
  family},}\ }\href {\doibase 10.1126/sciadv.1602241} {\bibfield  {journal}
  {\bibinfo  {journal} {Science Advances}\ }\textbf {\bibinfo {volume} {3}}
  (\bibinfo {year} {2017}{\natexlab{b}}),\ 10.1126/sciadv.1602241},\ \Eprint
  {http://arxiv.org/abs/http://advances.sciencemag.org/content/3/4/e1602241.full.pdf}
  {http://advances.sciencemag.org/content/3/4/e1602241.full.pdf} \BibitemShut
  {NoStop}%
\bibitem [{\citenamefont {Nguyen}\ \emph {et~al.}(2019)\citenamefont {Nguyen},
  \citenamefont {Pham}, \citenamefont {Nguyen}, \citenamefont {Nguyen},
  \citenamefont {Kino}, \citenamefont {Miyake},\ and\ \citenamefont
  {Dam}}]{nguyen2019regression}%
  \BibitemOpen
  \bibfield  {author} {\bibinfo {author} {\bibfnamefont {D.-N.}\ \bibnamefont
  {Nguyen}}, \bibinfo {author} {\bibfnamefont {T.-L.}\ \bibnamefont {Pham}},
  \bibinfo {author} {\bibfnamefont {V.-C.}\ \bibnamefont {Nguyen}}, \bibinfo
  {author} {\bibfnamefont {A.-T.}\ \bibnamefont {Nguyen}}, \bibinfo {author}
  {\bibfnamefont {H.}~\bibnamefont {Kino}}, \bibinfo {author} {\bibfnamefont
  {T.}~\bibnamefont {Miyake}}, \ and\ \bibinfo {author} {\bibfnamefont {H.-C.}\
  \bibnamefont {Dam}},\ }\bibfield  {title} {\enquote {\bibinfo {title} {A
  regression-based model evaluation of the curie temperature of
  transition-metal rare-earth compounds},}\ }in\ \href@noop {} {\emph {\bibinfo
  {booktitle} {Journal of Physics: Conference Series}}},\ Vol.\ \bibinfo
  {volume} {1290}\ (\bibinfo {organization} {IOP Publishing},\ \bibinfo {year}
  {2019})\ p.\ \bibinfo {pages} {012009}\BibitemShut {NoStop}%
\bibitem [{\citenamefont {Zhai}, \citenamefont {Chen},\ and\ \citenamefont
  {Lu}(2018)}]{zhai2018accelerated}%
  \BibitemOpen
  \bibfield  {author} {\bibinfo {author} {\bibfnamefont {X.}~\bibnamefont
  {Zhai}}, \bibinfo {author} {\bibfnamefont {M.}~\bibnamefont {Chen}}, \ and\
  \bibinfo {author} {\bibfnamefont {W.}~\bibnamefont {Lu}},\ }\bibfield
  {title} {\enquote {\bibinfo {title} {Accelerated search for perovskite
  materials with higher curie temperature based on the machine learning
  methods},}\ }\href@noop {} {\bibfield  {journal} {\bibinfo  {journal}
  {Computational Materials Science}\ }\textbf {\bibinfo {volume} {151}},\
  \bibinfo {pages} {41--48} (\bibinfo {year} {2018})}\BibitemShut {NoStop}%
\bibitem [{\citenamefont {Long}\ \emph {et~al.}(2021)\citenamefont {Long},
  \citenamefont {Fortunato}, \citenamefont {Zhang}, \citenamefont
  {Gutfleisch},\ and\ \citenamefont {Zhang}}]{long2021accelerating}%
  \BibitemOpen
  \bibfield  {author} {\bibinfo {author} {\bibfnamefont {T.}~\bibnamefont
  {Long}}, \bibinfo {author} {\bibfnamefont {N.~M.}\ \bibnamefont {Fortunato}},
  \bibinfo {author} {\bibfnamefont {Y.}~\bibnamefont {Zhang}}, \bibinfo
  {author} {\bibfnamefont {O.}~\bibnamefont {Gutfleisch}}, \ and\ \bibinfo
  {author} {\bibfnamefont {H.}~\bibnamefont {Zhang}},\ }\bibfield  {title}
  {\enquote {\bibinfo {title} {An accelerating approach of designing
  ferromagnetic materials via machine learning modeling of magnetic ground
  state and curie temperature},}\ }\href@noop {} {\bibfield  {journal}
  {\bibinfo  {journal} {Materials Research Letters}\ }\textbf {\bibinfo
  {volume} {9}},\ \bibinfo {pages} {169--174} (\bibinfo {year}
  {2021})}\BibitemShut {NoStop}%
\bibitem [{\citenamefont {Dam}\ \emph {et~al.}(2018)\citenamefont {Dam},
  \citenamefont {Nguyen}, \citenamefont {Pham}, \citenamefont {Nguyen},
  \citenamefont {Terakura}, \citenamefont {Miyake},\ and\ \citenamefont
  {Kino}}]{dam2018important}%
  \BibitemOpen
  \bibfield  {author} {\bibinfo {author} {\bibfnamefont {H.~C.}\ \bibnamefont
  {Dam}}, \bibinfo {author} {\bibfnamefont {V.~C.}\ \bibnamefont {Nguyen}},
  \bibinfo {author} {\bibfnamefont {T.~L.}\ \bibnamefont {Pham}}, \bibinfo
  {author} {\bibfnamefont {A.~T.}\ \bibnamefont {Nguyen}}, \bibinfo {author}
  {\bibfnamefont {K.}~\bibnamefont {Terakura}}, \bibinfo {author}
  {\bibfnamefont {T.}~\bibnamefont {Miyake}}, \ and\ \bibinfo {author}
  {\bibfnamefont {H.}~\bibnamefont {Kino}},\ }\bibfield  {title} {\enquote
  {\bibinfo {title} {Important descriptors and descriptor groups of curie
  temperatures of rare-earth transition-metal binary alloys},}\ }\href@noop {}
  {\bibfield  {journal} {\bibinfo  {journal} {Journal of the Physical Society
  of Japan}\ }\textbf {\bibinfo {volume} {87}},\ \bibinfo {pages} {113801}
  (\bibinfo {year} {2018})}\BibitemShut {NoStop}%
\bibitem [{\citenamefont {Hu}\ \emph {et~al.}(2020)\citenamefont {Hu},
  \citenamefont {Zhang}, \citenamefont {Fan}, \citenamefont {Li}, \citenamefont
  {Zhao}, \citenamefont {He}, \citenamefont {Zhao}, \citenamefont {Liu},\ and\
  \citenamefont {Xie}}]{Hu2020JPCM}%
  \BibitemOpen
  \bibfield  {author} {\bibinfo {author} {\bibfnamefont {X.}~\bibnamefont
  {Hu}}, \bibinfo {author} {\bibfnamefont {Y.}~\bibnamefont {Zhang}}, \bibinfo
  {author} {\bibfnamefont {S.}~\bibnamefont {Fan}}, \bibinfo {author}
  {\bibfnamefont {X.}~\bibnamefont {Li}}, \bibinfo {author} {\bibfnamefont
  {Z.}~\bibnamefont {Zhao}}, \bibinfo {author} {\bibfnamefont {C.}~\bibnamefont
  {He}}, \bibinfo {author} {\bibfnamefont {Y.}~\bibnamefont {Zhao}}, \bibinfo
  {author} {\bibfnamefont {Y.}~\bibnamefont {Liu}}, \ and\ \bibinfo {author}
  {\bibfnamefont {W.}~\bibnamefont {Xie}},\ }\bibfield  {title} {\enquote
  {\bibinfo {title} {Searching high spin polarization ferromagnet in heusler
  alloy via machine learning},}\ }\href {\doibase 10.1088/1361-648X/ab6e96}
  {\bibfield  {journal} {\bibinfo  {journal} {J. Phys.: Condens. Matter}\
  }\textbf {\bibinfo {volume} {32}},\ \bibinfo {pages} {205901} (\bibinfo
  {year} {2020})}\BibitemShut {NoStop}%
\bibitem [{\citenamefont {Lu}\ \emph {et~al.}(2022)\citenamefont {Lu},
  \citenamefont {Zhou}, \citenamefont {Guo},\ and\ \citenamefont
  {Wang}}]{Lu2022Chem}%
  \BibitemOpen
  \bibfield  {author} {\bibinfo {author} {\bibfnamefont {S.}~\bibnamefont
  {Lu}}, \bibinfo {author} {\bibfnamefont {Q.}~\bibnamefont {Zhou}}, \bibinfo
  {author} {\bibfnamefont {Y.}~\bibnamefont {Guo}}, \ and\ \bibinfo {author}
  {\bibfnamefont {J.}~\bibnamefont {Wang}},\ }\bibfield  {title} {\enquote
  {\bibinfo {title} {{On-the-fly interpretable machine learning for rapid
  discovery of two-dimensional ferromagnets with high Curie temperature}},}\
  }\href {\doibase 10.1016/j.chempr.2021.11.009} {\bibfield  {journal}
  {\bibinfo  {journal} {Chem}\ }\textbf {\bibinfo {volume} {8}},\ \bibinfo
  {pages} {769--783} (\bibinfo {year} {2022})}\BibitemShut {NoStop}%
\bibitem [{\citenamefont {Pedregosa}\ \emph {et~al.}(2011)\citenamefont
  {Pedregosa}, \citenamefont {Varoquaux}, \citenamefont {Gramfort},
  \citenamefont {Michel}, \citenamefont {Thirion}, \citenamefont {Grisel},
  \citenamefont {Blondel}, \citenamefont {Prettenhofer}, \citenamefont {Weiss},
  \citenamefont {Dubourg}, \citenamefont {Vanderplas}, \citenamefont {Passos},
  \citenamefont {Cournapeau}, \citenamefont {Brucher}, \citenamefont {Perrot},\
  and\ \citenamefont {Duchesnay}}]{tsne}%
  \BibitemOpen
  \bibfield  {author} {\bibinfo {author} {\bibfnamefont {F.}~\bibnamefont
  {Pedregosa}}, \bibinfo {author} {\bibfnamefont {G.}~\bibnamefont
  {Varoquaux}}, \bibinfo {author} {\bibfnamefont {A.}~\bibnamefont {Gramfort}},
  \bibinfo {author} {\bibfnamefont {V.}~\bibnamefont {Michel}}, \bibinfo
  {author} {\bibfnamefont {B.}~\bibnamefont {Thirion}}, \bibinfo {author}
  {\bibfnamefont {O.}~\bibnamefont {Grisel}}, \bibinfo {author} {\bibfnamefont
  {M.}~\bibnamefont {Blondel}}, \bibinfo {author} {\bibfnamefont
  {P.}~\bibnamefont {Prettenhofer}}, \bibinfo {author} {\bibfnamefont
  {R.}~\bibnamefont {Weiss}}, \bibinfo {author} {\bibfnamefont
  {V.}~\bibnamefont {Dubourg}}, \bibinfo {author} {\bibfnamefont
  {J.}~\bibnamefont {Vanderplas}}, \bibinfo {author} {\bibfnamefont
  {A.}~\bibnamefont {Passos}}, \bibinfo {author} {\bibfnamefont
  {D.}~\bibnamefont {Cournapeau}}, \bibinfo {author} {\bibfnamefont
  {M.}~\bibnamefont {Brucher}}, \bibinfo {author} {\bibfnamefont
  {M.}~\bibnamefont {Perrot}}, \ and\ \bibinfo {author} {\bibfnamefont
  {E.}~\bibnamefont {Duchesnay}},\ }\bibfield  {title} {\enquote {\bibinfo
  {title} {Scikit-learn: Machine learning in {P}ython},}\ }\href@noop {}
  {\bibfield  {journal} {\bibinfo  {journal} {Journal of Machine Learning
  Research}\ }\textbf {\bibinfo {volume} {12}},\ \bibinfo {pages} {2825--2830}
  (\bibinfo {year} {2011})}\BibitemShut {NoStop}%
\bibitem [{\citenamefont {Balachandran}(2019)}]{balachandran2019machine}%
  \BibitemOpen
  \bibfield  {author} {\bibinfo {author} {\bibfnamefont {P.~V.}\ \bibnamefont
  {Balachandran}},\ }\bibfield  {title} {\enquote {\bibinfo {title} {Machine
  learning guided design of functional materials with targeted properties},}\
  }\href@noop {} {\bibfield  {journal} {\bibinfo  {journal} {Computational
  Materials Science}\ }\textbf {\bibinfo {volume} {164}},\ \bibinfo {pages}
  {82--90} (\bibinfo {year} {2019})}\BibitemShut {NoStop}%
\bibitem [{\citenamefont {Balachandran}, \citenamefont {Xue},\ and\
  \citenamefont {Lookman}(2016)}]{balachandran2016structure}%
  \BibitemOpen
  \bibfield  {author} {\bibinfo {author} {\bibfnamefont {P.~V.}\ \bibnamefont
  {Balachandran}}, \bibinfo {author} {\bibfnamefont {D.}~\bibnamefont {Xue}}, \
  and\ \bibinfo {author} {\bibfnamefont {T.}~\bibnamefont {Lookman}},\
  }\bibfield  {title} {\enquote {\bibinfo {title} {Structure--curie temperature
  relationships in batio 3-based ferroelectric perovskites: anomalous behavior
  of (ba, cd) tio 3 from dft, statistical inference, and experiments},}\
  }\href@noop {} {\bibfield  {journal} {\bibinfo  {journal} {Physical Review
  B}\ }\textbf {\bibinfo {volume} {93}},\ \bibinfo {pages} {144111} (\bibinfo
  {year} {2016})}\BibitemShut {NoStop}%
\bibitem [{\citenamefont {Chen}\ \emph {et~al.}(2022)\citenamefont {Chen},
  \citenamefont {Lu}, \citenamefont {Wan}, \citenamefont {Chen}, \citenamefont
  {Zhou},\ and\ \citenamefont {Wang}}]{chen2022accurate}%
  \BibitemOpen
  \bibfield  {author} {\bibinfo {author} {\bibfnamefont {X.}~\bibnamefont
  {Chen}}, \bibinfo {author} {\bibfnamefont {S.}~\bibnamefont {Lu}}, \bibinfo
  {author} {\bibfnamefont {X.}~\bibnamefont {Wan}}, \bibinfo {author}
  {\bibfnamefont {Q.}~\bibnamefont {Chen}}, \bibinfo {author} {\bibfnamefont
  {Q.}~\bibnamefont {Zhou}}, \ and\ \bibinfo {author} {\bibfnamefont
  {J.}~\bibnamefont {Wang}},\ }\bibfield  {title} {\enquote {\bibinfo {title}
  {Accurate property prediction with interpretable machine learning model for
  small datasets via transformed atom vector},}\ }\href@noop {} {\bibfield
  {journal} {\bibinfo  {journal} {Physical Review Materials}\ }\textbf
  {\bibinfo {volume} {6}},\ \bibinfo {pages} {123803} (\bibinfo {year}
  {2022})}\BibitemShut {NoStop}%
\bibitem [{\citenamefont {Jacobs}\ \emph {et~al.}(2020)\citenamefont {Jacobs},
  \citenamefont {Mayeshiba}, \citenamefont {Afflerbach}, \citenamefont {Miles},
  \citenamefont {Williams}, \citenamefont {Turner}, \citenamefont {Finkel},\
  and\ \citenamefont {Morgan}}]{mastml}%
  \BibitemOpen
  \bibfield  {author} {\bibinfo {author} {\bibfnamefont {R.}~\bibnamefont
  {Jacobs}}, \bibinfo {author} {\bibfnamefont {T.}~\bibnamefont {Mayeshiba}},
  \bibinfo {author} {\bibfnamefont {B.}~\bibnamefont {Afflerbach}}, \bibinfo
  {author} {\bibfnamefont {L.}~\bibnamefont {Miles}}, \bibinfo {author}
  {\bibfnamefont {M.}~\bibnamefont {Williams}}, \bibinfo {author}
  {\bibfnamefont {M.}~\bibnamefont {Turner}}, \bibinfo {author} {\bibfnamefont
  {R.}~\bibnamefont {Finkel}}, \ and\ \bibinfo {author} {\bibfnamefont
  {D.}~\bibnamefont {Morgan}},\ }\bibfield  {title} {\enquote {\bibinfo {title}
  {The materials simulation toolkit for machine learning (mast-ml): An
  automated open source toolkit to accelerate data-driven materials
  research},}\ }\href@noop {} {\bibfield  {journal} {\bibinfo  {journal}
  {Computational Materials Science}\ }\textbf {\bibinfo {volume} {176}},\
  \bibinfo {pages} {109544} (\bibinfo {year} {2020})}\BibitemShut {NoStop}%
\bibitem [{\citenamefont {Murdock}\ \emph {et~al.}(2020)\citenamefont
  {Murdock}, \citenamefont {Kauwe}, \citenamefont {Wang},\ and\ \citenamefont
  {Sparks}}]{murdock2020domain}%
  \BibitemOpen
  \bibfield  {author} {\bibinfo {author} {\bibfnamefont {R.~J.}\ \bibnamefont
  {Murdock}}, \bibinfo {author} {\bibfnamefont {S.~K.}\ \bibnamefont {Kauwe}},
  \bibinfo {author} {\bibfnamefont {A.~Y.-T.}\ \bibnamefont {Wang}}, \ and\
  \bibinfo {author} {\bibfnamefont {T.~D.}\ \bibnamefont {Sparks}},\ }\bibfield
   {title} {\enquote {\bibinfo {title} {Is domain knowledge necessary for
  machine learning materials properties?}}\ }\href@noop {} {\bibfield
  {journal} {\bibinfo  {journal} {Integrating Materials and Manufacturing
  Innovation}\ }\textbf {\bibinfo {volume} {9}},\ \bibinfo {pages} {221--227}
  (\bibinfo {year} {2020})}\BibitemShut {NoStop}%
\bibitem [{\citenamefont {Marc}\ \emph {et~al.}(2019)\citenamefont {Marc},
  \citenamefont {Weinstein}, \citenamefont {tgwoodcock}, \citenamefont {Simon},
  \citenamefont {chebee7i}, \citenamefont {Morgan}, \citenamefont {Knight},
  \citenamefont {Swanson-Hysell}, \citenamefont {Evans}, \citenamefont
  {jl~bernal}, \citenamefont {ZGainsforth}, \citenamefont {Badger},
  \citenamefont {SaxonAnglo}, \citenamefont {Greco},\ and\ \citenamefont
  {Zuidhof}}]{pythonternary}%
  \BibitemOpen
  \bibfield  {author} {\bibinfo {author} {\bibnamefont {Marc}}, \bibinfo
  {author} {\bibfnamefont {B.}~\bibnamefont {Weinstein}}, \bibinfo {author}
  {\bibnamefont {tgwoodcock}}, \bibinfo {author} {\bibfnamefont
  {C.}~\bibnamefont {Simon}}, \bibinfo {author} {\bibnamefont {chebee7i}},
  \bibinfo {author} {\bibfnamefont {W.}~\bibnamefont {Morgan}}, \bibinfo
  {author} {\bibfnamefont {V.}~\bibnamefont {Knight}}, \bibinfo {author}
  {\bibfnamefont {N.}~\bibnamefont {Swanson-Hysell}}, \bibinfo {author}
  {\bibfnamefont {M.}~\bibnamefont {Evans}}, \bibinfo {author} {\bibnamefont
  {jl~bernal}}, \bibinfo {author} {\bibnamefont {ZGainsforth}}, \bibinfo
  {author} {\bibfnamefont {T.~G.}\ \bibnamefont {Badger}}, \bibinfo {author}
  {\bibnamefont {SaxonAnglo}}, \bibinfo {author} {\bibfnamefont
  {M.}~\bibnamefont {Greco}}, \ and\ \bibinfo {author} {\bibfnamefont
  {G.}~\bibnamefont {Zuidhof}},\ }\href {\doibase 10.5281/zenodo.2628066}
  {\enquote {\bibinfo {title} {marcharper/python-ternary: Version 1.0.6},}\ }
  (\bibinfo {year} {2019})\BibitemShut {NoStop}%
\bibitem [{\citenamefont {{National Minerals Information
  Center}}(2023)}]{sciencebase}%
  \BibitemOpen
  \bibfield  {author} {\bibinfo {author} {\bibnamefont {{National Minerals
  Information Center}}},\ }\href {\doibase 10.5066/P9WCYUI6} {\enquote
  {\bibinfo {title} {U.s. geological survey mineral commodity summaries 2023
  data release},}\ } (\bibinfo {year} {2023})\BibitemShut {NoStop}%
\end{thebibliography}%

\end{document}


\title{Machine Learning Predictions of High-Curie-Temperature Materials}
\maketitle

\begin{figure*}
    \begin{minipage}[b]{0.45\linewidth}
        \centering
        \includegraphics[width=\textwidth]{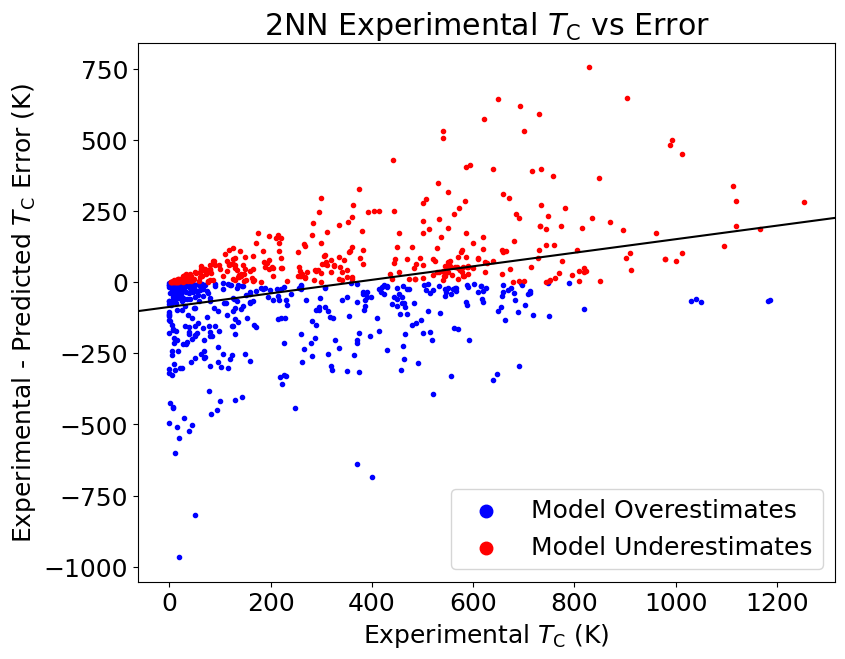}
        \label{fig:figure1}
    \end{minipage}
    \hspace{0.1cm}
    \begin{minipage}[b]{0.45\linewidth}
        \centering
        \includegraphics[width=\textwidth]{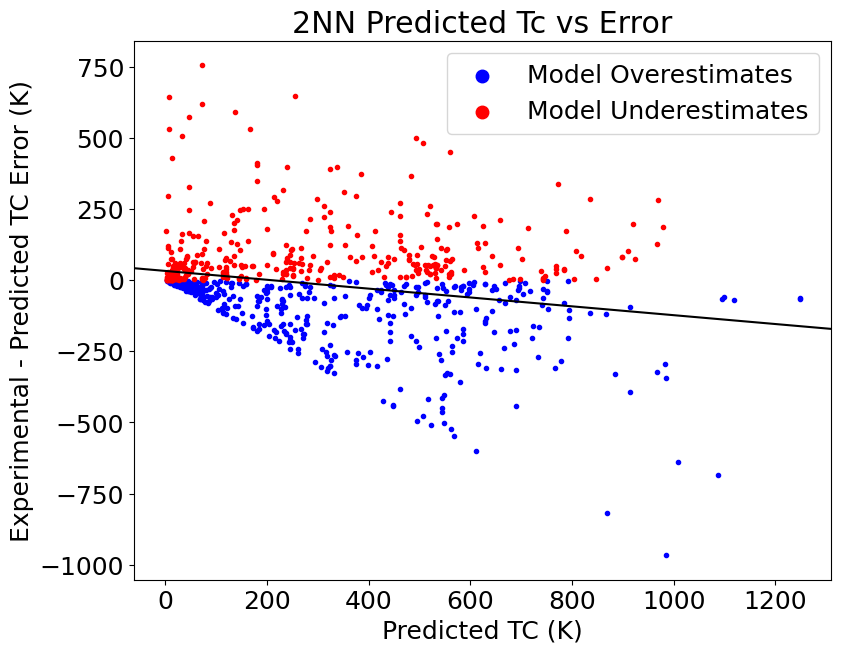}
        \label{fig:figure2}
    \end{minipage}
    \caption{Prediction error produced from the two-nearest neighbor model for DS1. The error is the predicted Curie temperature subtracted from the experimental Curie temperature.(Left) Error plotted against the experimental $T_\mathrm{C}$ values. (Right) Error plotted against the predicted $T_\mathrm{C}$ values. The black line is the line of best fit.}
    \label{fig:KNN error}
\end{figure*}

\begin{figure}
\caption{Random forest mean absolute error with principal component analysis (PCA). The MAE of the random forest prediction was found using training data containing 5 up to 85 features. The features were reduced from the original 85 using the PCA method.}
\includegraphics[width=8cm]{figures/Main_Paper/PCA_Random Forest_-_DS1.png}
\label{fig:pca}
\end{figure}